\documentclass[12pt,preprint]{aastex}

\shorttitle{CO Molecular Gas}
\shortauthors{Yao et al.}

\begin{document}


\title{CO Molecular Gas in Infrared Luminous Galaxies}


\author{Lihong Yao  and E. R. Seaquist}
\affil{Department of Astronomy and Astrophysics, University of Toronto,
    Toronto, Ontario M5S 3H8, Canada}
\email{yao@astro.utoronto.ca, seaquist@astro.utoronto.ca}

\author{Nario Kuno}
\affil{Nobeyama Radio Observatory, Minamimaki, Minamisaku, Nagano, 384-1305, Japan}
\email{kuno@nro.nao.ac.jp}

\and

\author{Loretta Dunne}
\affil{Department of Physics and Astronomy, University of Wales, Cardiff, CF24 3YB, UK}
\email{Loretta.Dunne@astro.cf.ac.uk}




\begin{abstract}
We present the first statistical survey of the properties of the $^{12}$CO(1-0) and 
$^{12}$CO(3-2) line emission from the nuclei of a nearly complete subsample of 60 
infrared (IR) luminous galaxies selected from SCUBA Local Universe Galaxy Survey 
(SLUGS). This subsample is flux limited at $S_{60 \mu m}$ $\ge$ 5.24 Jy with far-IR 
(FIR) luminosities mostly at $L_{FIR}$ $>$ 10$^{10}$ L$_{\odot}$. We compare the 
emission line strengths of $^{12}$CO(1-0) and (3-2) transitions at a common resolution 
of 15$^{\prime\prime}$. The measured $^{12}$CO(3-2) to (1-0) line intensity ratios 
$r_{31}$ vary from 0.22 to 1.72 with a mean value of 0.66 for the sources observed,
indicating a large spread of the degree of excitation of CO in the sample. These CO
data, together with a wide range of data at different wavelengths obtained from the 
literature, allow us to study the relationship between the CO excitation conditions
and the physical properties of gas/dust and star formation in the central regions
of galaxies. Our analysis shows that there is a non-linear relation between
CO and FIR luminosities, such that their ratio $L_{CO}$/$L_{FIR}$ decreases linearly
with increasing $L_{FIR}$. This behavior was found to be consistent with 
the Schmidt Law relating star formation rate to molecular gas content, with an index 
$N$ = 1.4 $\pm$ 0.3. We also find a possible dependence of the degree of CO gas 
excitation on the efficiency of star forming activity. Using the large velocity 
gradient (LVG) approximation to model the observed data, we investigate the CO-to-H$_2$ 
conversion factor $X$ for the SLUGS sample. The results show that the mean value of 
$X$ for the SLUGS sample is lower by a factor of 10 compared to the conventional 
value derived for the Galaxy, if we assume the abundance of CO relative to H$_2$, 
$Z_{CO}$ = 10$^{-4}$. For a subset of 12 galaxies with H I maps, we derive a mean 
total face-on surface density of H$_2$+H I of about 42 M$_{\odot}$ pc$^{-2}$ within 
about 2 kpc of the nucleus. This value is intermediate between that in galaxies like 
our own and those with strong star formation.  
  
\end{abstract}




\section{Introduction} \label{intro}

Knowledge of the properties and evolution of the gas and dust content in 
IR luminous galaxies is essential for understanding the cause and temporal 
evolution of starburst activity. In particular, studies of such galaxies 
in the nearby universe are essential as a step in understanding the role 
of the starburst phenomenon in the cosmic evolution of galaxies. Therefore, 
there is a need to investigate large statistical samples of IR luminous 
galaxies using a multitude of different types of data, including CO, H I, and 
continuum in the submillimeter (sub-mm), FIR and radio, in order to constrain 
theories of how the interstellar medium (ISM) evolves \citep[e.g.][]{ene96, ene97}. 
A significant step was achieved by Dunne et al. (2000) for the nearby universe 
by investigating the sub-mm properties of the dust in a complete sample of
104 galaxies. The SLUGS survey is based on the Revised Bright Galaxy Sample of 
{\it IRAS} galaxies \citep{soi87} within -10$^{\circ}$ $\le$ $\delta$ $\le$ 
+50$^{\circ}$ and with $cz$ $\ge$ 1900 km s$^{-1}$ and a flux limit of 
$S_{60\mu m}$ $\ge$ 5.24 Jy. By virtue of the selection, the SLUGS galaxies 
have the FIR luminosity $L_{FIR}$ $\ge$ 10$^{10}$ L$_{\odot}$, and hence 
have greater than the average star formation rate. A study of molecular gas of 
a complete flux-limited subsample is important to complement the study of the dust. 

Low lying CO rotational line transitions at mm and sub-mm wavelengths are often 
used as tracers of molecular hydrogen. It is valid for molecular clouds in the 
Galaxy and also for external galaxies. The J = 3 level is at 33 K with a critical 
density of 1.5 $\times$ 10$^4$ cm$^{-3}$, compared to the J = 1 and 2 levels of 
$^{12}$CO are lying respectively at 5.5 and 17 K above the ground level, and with 
respective critical densities 7.4 $\times$ 10$^2$ and 4.9 $\times$ 10$^3$ cm$^{-3}$
for a kinetic temperature $T_{kin}$ = 30 K in the optically thin limit \citep{jan95}. 
Therefore, the $^{12}$CO(3-2) transition is a better tracer for the warmer and/or 
denser molecular gas in star forming regions. Furthermore, the ratio of $^{12}$CO(3-2) 
to (1-0) line emission provides a more sensitive measure of the gas temperature and 
density than the ratio of $^{12}$CO(2-1) to (1-0) lines. The first detection of the 
$^{12}$CO(3-2) line in the nearby starburst galaxy was that in IC 342 reported by 
Ho (1987). Many observations of the $^{12}$CO(1-0) and $^{12}$CO(2-1) lines have 
since been made in local galaxies \citep[e.g.][and many others]{san91, bra92, bra93, 
you95, chi96}. Most observations of the $^{12}$CO(3-2) line cover the central region 
of nearby objects \citep[e.g.][]{dev94, mau99, dum01}, where the physical conditions 
of molecular gas may be different from those prevailing in molecular clouds in the disk 
of the Galaxy. Extended maps of $^{12}$CO(3-2) line for three nearby normal galaxies 
were reported by Wielebinski et al. (1999). 

The weaknesses of most of the previous studies on CO molecular gas are the small 
numbers of objects or the lack of uniform selection criteria, whereas their strength 
is the production of maps with high linear resolution. In this paper, we 
address the aforementioned weaknesses by presenting the largest statistical survey 
of $^{12}$CO(1-0) and $^{12}$CO(3-2) point measurements from the central regions 
in a complete subsample of 60 IR luminous sources from SLUGS \citep{dun00}. The 
angular resolutions of the $^{12}$CO(1-0) observations at the Nobeyama Radio Observatory 
(NRO) and the $^{12}$CO(3-2) observations at the James Clerk Maxwell Telescope (JCMT) 
are nearly identical, so that no assumptions need be made about differing beam 
dilution factors. The beamsize ($\sim$ 15$^\prime$$^\prime$) is also nearly 
identical to that of the SLUGS survey by Dunne et al. (2000). Together with 
the 850 $\mu$m fluxes, plus existing data on H I \citep{tho02}, radio continuum 
\citep{con90}, and far-IR \citep[][NED\footnote{The NASA/IPAC Extragalactic Database 
(NED) is operated by the Jet propulsion Laboratory, California Institute of Technology, 
under contract with the National Aeronautics and Space Administration.}]{dun00},
we are able to search for a relationship between the degree of excitation of the CO 
in this sample and the star forming properties. 

It is now suspected that the $^{12}$CO(3-2) line emission makes up a substantial 
fraction of the SCUBA flux at 850 $\mu$m for starburst galaxies \citep[e.g.][]{pna00, 
zhu03}. Our $^{12}$CO(3-2) observations provide an important database for 
correcting the SCUBA sub-mm data in the SLUGS sample, thus permitting a revised 
characterization of the dust, which will be presented in a separate paper. 

Finally, our investigations on the gas and dust properties in starburst regions 
provide an opportunity to investigate the controversial parameter $X$, the ratio 
of molecular hydrogen mass $M$(H$_2$) and the $^{12}$CO(1-0) luminosity $L_{CO}$, 
based on the LVG method \citep{gnk74}. We obtain an estimate for this parameter 
which depends on $Z_{CO}$, the abundance of CO relative to H$_2$. We also use our 
estimate of this parameter to investigate the molecular gas content of our SLUGS 
subsample.

We assume $q_0$ = 0.5 and $H_0$ = 75 km s$^{-1}$ Mpc$^{-1}$ throughout for consistency 
with previous work on these galaxies \citep[e.g.][]{dun00}.

\section{Observations} \label{obs}

\subsection{$^{12}$CO(1-0) Observations at NRO} \label{nrotel}

Two observing runs measuring the $^{12}$CO(1-0) line at 115.27 GHz for 46 SLUGS 
galaxies were made with NRO 45m telescope between 2001 April 12 - 19 and 2001 
November 28 - December 1, using the SIS receiver S100 (single side band or SSB). 
The wide band acousto-optical spectrometer (AOS) was used as the back end which 
provided a total usable bandwidth of 250 MHz and a frequency resolution of 250 KHz. 
At 115.27 GHz, these correspond to a total velocity coverage of 650 km s$^{-1}$ 
and velocity resolution of 0.65 km s$^{-1}$. The NRO 45m telescope beamsize (FWHM) 
at this frequency is 14.6$^\prime$$^\prime$. The standard source Mars was used 
for the determination of the main beam efficiency. The measurement yields $\eta_{mb}$ 
= 0.34 $\pm$ 0.04 after averaging the results from the first and second runs. 
The average uncertainty in the calibration of spectral line is $\sim$ 10\%.  

Position switching in azimuth was employed with about 20 seconds each for both ON and 
OFF positions, 10 sec to move from ON to OFF or from OFF back to ON, so that about one 
minute was required for each ON-OFF sequence or scan. Since about 30 scans were
made for each source, the total integration time per pointing is about 30 minutes. 
Telescope pointing was checked every half hour by observing strong SiO maser lines 
from nearby evolved stars with receiver S40 at 43 GHz. The average pointing error 
was $\sim$ 3.0$^\prime$$^\prime$ (r.m.s.) for the first observing run and $\sim$ 
2.6$^\prime$$^\prime$ (r.m.s.) for the second run at low wind speed. The weather 
conditions were average during the NRO observing runs and the system temperature 
for good weather conditions was $T_{sys}$ $\sim$ 535 K. However, half of the observing 
time was lost for each run due to poor weather conditions. The total number of sources 
observed at this transition is 51, out of a total sample of 60.

\subsection{$^{12}$CO(3-2) Observations at JCMT} \label{jcmttel}

Observations of the $^{12}$CO(3-2) line at 345.80 GHz for the selected 60 SLUGS 
galaxies were obtained with the 15m JCMT on 2001 April 23 - 28 and 2002 June 25. 
We used the dual-channel receiver B3 in single-sideband mode, and the Digital 
Autocorrelation Spectrometer (DAS) spectrometer with a bandwidth of 920 MHz 
($\sim$ 800 km s$^{-1}$) and a frequency resolution of 750 KHz ($\sim$ 0.65 km s$^{-1}$).
Focus and pointing were monitored frequently by observing bright continuum 
sources and planets. The average pointing error (r.m.s.) was found to be $\sim$ 
3.2$^\prime$$^\prime$. The observations were performed by using beam-switching at a 
frequency of 1 Hz with a beam throw of 180$^\prime$$^\prime$ in azimuth. The JCMT 
beamsize (FWHM) at this frequency is 14.4$^\prime$$^\prime$. The main-beam 
efficiency was determined by averaging four measurements based on observations of Mars,
yielding a value of $\eta_{mb}$ = 0.50 $\pm$ 0.05. The spectral line calibration 
was checked each night by measuring $^{12}$CO(3-2) line emission from IRC+10216.
The mean calibration error is $\sim$ 10\%. The weather conditions for these 
observations were good, and the average system temperature was $T_{sys}$ $\sim$ 600 K. 

\section{Results}

\subsection{The Spectra}

The reduction of the $^{12}$CO(1-0) data was performed using the NRO NEWSTAR 
package, and the $^{12}$CO J=3-2 data were reduced using the JCMT SPECX package.
The $^{12}$CO(1-0) line spectra for 51 galaxies together with the spectra of 
$^{12}$CO(3-2) line emission for 59 galaxies are presented in Fig.~\ref{cospectra}. 
A linear baseline was subtracted from each spectrum for both data sets, and the spectra
were smoothed to 10.2 km s$^{-1}$ and 11.2 km s$^{-1}$ for the NRO and the JCMT data 
respectively. We converted the measured antenna temperature scale $T^{\ast}_{a}$ of 
both spectra to the main-beam temperature $T_{mb}$ scale by using the relation 
$T_{mb}$ = $T^{\ast}_{a}$/$\eta_{mb}$. The main-beam efficiencies of NRO and 
JCMT telescopes are given in \S~\ref{obs}. The JCMT and NRO line profile shapes 
mostly agree with each other. The ones that show discrepancies may in some instances 
be due to pointing error.

\subsection{CO and FIR Luminosities} \label{Lum}

The velocity-integrated intensities of $^{12}$CO(1-0) line emission $I_{10}$ and 
$^{12}$CO(3-2) line emission $I_{32}$ corresponding to the common telescope
beamsize of $\sim$ 15$^{\prime\prime}$ are shown in Table~\ref{tbl-1}. The 
uncertainties for the integrated line intensities include components due to noise, 
the uncertainty in the beam efficiencies and the calibration of spectral line, 
combined quadratically. The luminosities associated with the beamsize of 
$\sim$ 15$^{\prime\prime}$, in units K km s$^{-1}$ Mpc$^2$ $\Omega_{b}$, were 
calculated using equations 1(a) and 1(b) below and are presented in Table~\ref{tbl-2}. 
Both equations 1(a) and 1(b) were derived by transforming the relevant parameters 
from the rest frame to the observer frame.

\begin{mathletters}
\begin{eqnarray}
L_{^{12}CO(1-0)} & = & \Big[\frac{D^2_{L}}{(1+z)^3}\Big] I_{10} \Omega_{b} , \\
L_{^{12}CO(3-2)} & = & \Big[\frac{D^2_{L}}{(1+z)^3}\Big] I_{32} \Omega_{b}
\end{eqnarray}
\end{mathletters}

where $D_{L}$ is the luminosity distance of a galaxy in Mpc, $D_{L}$ = $D$ (1+z),
and $D$ is the proper distance of the galaxy (see Table~\ref{tbl-1}), 
the integrated line intensities $I_{10}$ and $I_{32}$ have units of K km s$^{-1}$, 
and $\Omega_{b}$ is the beam solid angle. The numerical value of $\Omega_{b}$ is 
constant for all observations and has been set to unity for simplicity. For conversion 
to absolute luminosities, the value of $\Omega_{b}$ = 1.1 $\times$ 10$^{-3}$ sterradians.
 
Fig.~\ref{LCO2Lfir} shows a comparison between the CO luminosities $L_{CO}$ and
FIR luminosities $L_{FIR}$ for two different FIR color regimes. The quantity $L_{FIR}$ 
here corresponds to the volume marked by a 15$^{\prime\prime}$ beam obtained by 
applying a scale factor to the total FIR luminosity (from Dunne et al. 2000). This 
factor is derived from the original SCUBA images which have a resolution of 
15$^{\prime\prime}$ and from radio continuum images \citep{con90} convolved to 
15$^{\prime\prime}$. It measures the ratio of flux within the beam to the total 
flux. It is less than unity if the source is resolved by the telescope beam and equal 
to unity for an unresolved source. It is assumed here that the FIR brightness 
distribution is similar to that for the sub-mm/cm and radio continuum. A linear fit 
to each transition in the log-log plane yields

\begin{mathletters}
\begin{eqnarray}
log_{10} L_{^{12}CO(1-0)} & = & (-0.9 \pm 0.7) + (0.58 \pm 0.07) log_{10} L_{FIR} , \\
log_{10} L_{^{12}CO(3-2)} & = & (-2.4 \pm 0.5) + (0.70 \pm 0.04) log_{10} L_{FIR}   
\end{eqnarray}
\end{mathletters}

The r.m.s. dispersion for the fits is about a factor of 3 and 2 for $^{12}$CO(1-0)
and $^{12}$CO(3-2) respectively. The scatter is much larger than the errors measured 
for the telescopes indicated by the error bars in the plots. Variations in the 
CO-to-H$_2$ conversion factor $X$ may be responsible for part of the scatter. The 
larger scatter seen in the $^{12}$CO(1-0) plot may be due to the larger baseline 
ripple measured at this transition. 

It is evident that the CO luminosities at both transitions increase non-linearly 
with increasing $L_{FIR}$, such that the ratio $L_{CO}$/$L_{FIR}$ decreases with 
increasing $L_{FIR}$. Fig.~\ref{rL2l} shows a plot of the ratio of $L_{CO}$/$L_{FIR}$ 
versus the projected beamsize on the galaxies, to test whether this effect is introduced 
by partially resolving the nearby galaxies. No relationship is evident, indicating 
that the total luminosity of the galaxy is the primary factor. The implications of 
this result will be discussed in \S~\ref{rL}.

\subsection{The $^{12}$CO(3-2)/(1-0) Line Intensity Ratio} \label{r31}

For the 48 galaxies with data at both transitions, we calculate the intensity 
ratios of the two transitions using the equation, 

\begin{equation}
r_{31} = I_{32} / I_{10} 
\end{equation}

The results along with the asymmetric uncertainties are given in Table~\ref{tbl-1}. 
The asymmetric errors of $r_{31}$ are derived analytically, assuming the error 
distributions of $I_{10}$ and $I_{32}$ are Gaussian functions. The asymmetric 
confidence intervals in Table~\ref{tbl-1} represent an enclosed probability of 68\%.

The values of $r_{31}$ range from 0.22 (for NGC 5665) to 1.72 (for NGC 5937). 
There are seven sources with $r_{31}$ $>$ 1, namely UGC 5376, IR 1222-06, UGC 8739, 
NGC 5433, NGC 5713, NGC 5937, and MCG +01-42, signifying very high excitation coupled 
with low optical depth. However, UGC 5376, IR 1222-06, and MCG +01-42 have high 
uncertainties in $r_{31}$, leaving only four whose $r_{31}$ values are significantly 
greater than 1.0. In our sample, thirteen galaxies have $r_{31}$ ratios less than 0.4, 
the average value of $r_{31}$ derived from the giant molecular clouds in the Galaxy 
\citep{san93}. 

The mean value of $r_{31}$ for SLUGS sample is 0.66, close to that of 0.63 measured 
for 28 galaxies with strong CO emission published by Mauersberger et al. (1999). 
The quantity $L_{FIR}$ ranges from 10$^9$ to 10$^{12}$ L$_{\odot}$ in the Mauersberger 
sample, compared to the range 10$^{10}$ - 10$^{12}$ L$_{\odot}$ in SLUGS sample. 
Fig.~\ref{r31his} shows a comparison between the distributions of $r_{31}$ for the 
SLUGS sample and those measured by Mauersberger et al. (1999). The probability of 
a significant difference between these two distributions is $\sim$ 68\% using a 
{\it Kolmogorov-Smirnov} test. This implies that these two distributions are not 
significantly different. The galaxies observed by Mauersberger et al. (1999) are 
located between 0.8 and 73 Mpc, much closer, and on average less luminous than our 
objects in the range 25 - 277 Mpc. Thus there is no evidence that $r_{31}$ depends on 
the distance of the galaxies over the range 0.8 - 277 Mpc, because of the different 
luminosity ranges represented by these two samples.  

A search was conducted for relationships between $r_{31}$ and a number of 
parameters relating primarily to star formation. The existence of such correlations
would be expected to shed light on the factors affecting star formation, or 
the effects of star formation on gas properties. These are considered below:

\subsubsection{Projected Beamsize} \label{pb}

Radio continuum maps \citep{con90}, the H I \citep{tho02}, and SCUBA 850$\mu$m maps 
\citep{tho02} show that much of the emission in the SLUGS objects is extended with 
respect to the beamsize of $\sim$ 15$^{\prime\prime}$, especially for the nearer
objects. On average we detect about 45\% of all the emission from these galaxies
assuming the CO is distributed in a similar way to the 850 $\mu$m and radio continuum
emission. Therefore, the CO measurements may reflect the physical conditions in the 
center regions for the galaxies at lower $cz$. Maps of $^{12}$CO(3-2) line emission 
in 12 nearby galaxies by Dumke et al. (2001) have shown that the $^{12}$CO(3-2)/(1-0) 
line intensity ratio decreases outward from the centers of many galaxies. In principle, 
for a given beamsize, the average line ratios obtained from the single pixel 
measurement should therefore decrease with increasing distance, since more 
emission from lesser excited CO lines arising from the disk regions is detected
by the telescope beam. Fig.~\ref{r2l} shows a plot of $r_{31}$ versus projected
beam size. Similar to the relation between the ratio of $L_{CO}/L_{FIR}$ and the
linear size sampled (see Fig.~\ref{rL2l}), no correlation between these quantities 
is evident. The lack of such a correlation may be due to sample selection effect, 
however, since the more distant galaxies may have a larger fraction of highly excited 
CO molecular gas.

\subsubsection{Dust and Gas} \label{DnG}

Large amounts of FIR emission are generated by the dust heated by young massive 
stars in starburst systems, in which case the FIR luminosity is a good measurement 
of the star formation rate ($SFR$) \citep{ken98}. Fig.~\ref{r2TdLfir} shows the 
$^{12}$CO line ratio versus dust temperature $T_{dust}$ and FIR luminosity 
$L_{FIR}$ within a 15$^{\prime\prime}$ beam. The dust temperatures are from Dunne 
et al. (2000) and the FIR luminosities are scaled to a 15$^{\prime\prime}$ beam using 
sub-mm and radio maps as described in \S~\ref{Lum}. No significant correlation is evident
in these plots. Other $SFR$ indicators are radio continuum luminosity $L_{radio}$ and 
the optical color index ($B - V$). Plots for these are not shown, but no obvious 
correlation was found with the parameter $r_{31}$. In Fig.~\ref{r2M}, we examine the 
dependence of $r_{31}$ on dust mass $M_{dust}$ and molecular hydrogen mass $M$(H$_2$). 
The dust masses were obtained from Dunne et al. (2000) and scaled to a 
15$^{\prime\prime}$ beam as described earlier. The molecular gas masses for the 
SLUGS sample are derived from our CO luminosities by applying the conversion factor 
$X$ = 2.7 $\times$ 10$^{19}$ cm$^{-2}$ [K km s$^{-1}$]$^{-1}$ derived in \S~\ref{X}. 
The result is $M$(H$_2$) = 1.1 $\times$ 10$^3$ $D^2_{L}$/(1+$z$) $S_{CO}$ M$_{\odot}$ 
by scaling the result of Kenney \& Young (1989), where $D_{L}$ is the luminosity 
distance of a galaxy in Mpc, and $S_{CO}$ is the $^{12}$CO(1-0) flux measured within 
a 15$^{\prime\prime}$ beam, $S_{CO}$ = 2.1 $I_{10}$ Jy km s$^{-1}$. The values of 
$M$(H$_2$) are also presented in Table~\ref{tbl-2}. Again, no relationship is evident.

\subsubsection{Surface Brightnesses and Color Indices}

The average surface brightness for all wavebands is defined here as the ratio of 
the total flux to the square of the optical diameter. Thus it measures the surface 
brightness averaged over the optical diameter, and provides a measure of the 
concentration of star forming activity. Fig.~\ref{r2sb} shows a plot of $r_{31}$
versus the surface brightness for mid-IR, far-IR, sub-mm, and H I emission. 
We find two possible correlations, one is between $r_{31}$ and FIR surface brightness, 
and another one is between the $r_{31}$ and SCUBA 850 $\mu$m surface brightness. 
Fig.~\ref{r2cidx} shows the line ratio versus different color indices
or ratios: $\Big[\frac{S_{60 \mu m}}{S_{100 \mu m}}\Big]$ and
$\Big[\frac{S_{H I}}{S_{850 \mu m}}\Big]$. No significant correlation is found  
between these ratios.

\subsubsection{Star Formation Efficiency}

The ratio $L_{FIR}$/$M$(H$_2$) is often used as an indicator of the star formation
efficiency ($SFE$) which gives the star formation rate per unit mass of molecular 
gas. In Table~\ref{tbl-2}, we present the results of $SFE$ calculations for our SLUGS 
sample. Here the mass of molecular gas has been derived in the same manner as that 
in \S~\ref{DnG}. These values are an order of magnitude higher than what would be 
obtained if the conventional value of $X$ is used. Fig.~\ref{r2sfe} shows a 
significant correlation between $r_{31}$ and the $SFE$ within the telescope beam.
The sample shown in Fig.~\ref{r2sfe} is divided into two ranges by gas mass centered 
at $M$(H$_2$) = 10$^8$ M$_{\odot}$, and also by dust FIR luminosity centered at 
$L_{FIR}$ = 10$^{10}$ L$_{\odot}$ which are indicated by four different symbols 
in the plot. The linear correlation coefficient is obtained from the data with 
$SFE$ $\le$ 200 L$_{\odot}$/M$_{\odot}$. The coefficient is 0.74 at a significance 
level of 1.4 $\times$ 10$^{-7}$ (i.e. probability that $r_{31}$ and $SFE$ are 
unrelated). However, the correlation is diminished at 
$SFE$ $>$ 200 L$_{\odot}$/M$_{\odot}$, where the ratios spread between 0.5 and 1.72. 
The significance of the dependence of $r_{31}$ on $SFE$ will be discussed 
in \S~\ref{corrl}. 

\section{Discussion} 

\subsection{The Relation Between CO and FIR Luminosities} \label{rL}

As noted in \S~\ref{Lum}, the non-linear variation of CO with FIR luminosity 
implies a decrease in the ratio $L_{CO}$/$L_{FIR}$ with increasing $L_{FIR}$ 
(see Fig.~\ref{LCO2Lfir}). For $^{12}$CO(1-0) this ratio decreases by more than an 
order of magnitude over the range $L_{FIR}$ = 5 $\times$ 10$^9$ - 10$^{12}$ L$_{\odot}$. 
Part of this variation might be attributable to the effect of beamsize if the ratio is 
dependent on the fraction of the galaxy resolved by the beam. However, as noted 
in \S~\ref{Lum}, Fig.~\ref{rL2l} shows that the beam effects do not seem to influence
the $L_{CO}$/$L_{FIR}$ ratio, and that for any plausible gas excitation gradient 
(i.e. hotter gas confined in the center) such effects would show in the $r_{31}$ ratio 
as well, but Fig.~\ref{r2l} shows no evidence that more distant objects have smaller 
values of $r_{31}$ (see \S~\ref{pb}). In addition, this effect would be expected to
produce an increase rather than a decrease in $L_{CO}$/$L_{FIR}$ with increasing 
$L_{FIR}$ because of the strong sensitivity of $L_{FIR}$ to dust temperature. 
Therefore this ratio depends intrinsically on the total FIR luminosity. It should 
be noted that our results imply that for high $z$ galaxies where $L_{FIR}$ is 
exceedingly high, the corresponding values of $L_{CO}$ will be seriously 
overestimated if a linear relation between $L_{CO}$ and $L_{FIR}$ were assumed. 

A possible cause of the variation in the $L_{CO}$/$L_{FIR}$ ratio is an increase in
$L_{FIR}$ per unit mass of dust with increasing $L_{FIR}$. To investigate this point 
further, we first plot in Fig.~\ref{LCO2Md} $L_{CO}$ versus $M_{dust}$ to determine 
this relationship is linear. Linear fits to these relations in the log-log plane 
yield

\begin{mathletters}
\begin{eqnarray}
log_{10} L_{^{12}CO(1-0)} & = & (-0.4 \pm 0.6) + (0.77 \pm 0.09) log_{10} M_{dust} , \\
log_{10} L_{^{12}CO(3-2)} & = & (-1.8 \pm 0.5) + (0.95 \pm 0.07) log_{10} M_{dust}    
\end{eqnarray}
\end{mathletters}

We see indeed that the slopes are closer to unity than with log$_{10}L_{CO}$ versus 
log$_{10}L_{FIR}$ (Equation (2a) and (2b)), though there is still a significant 
departures from a linear relation for $L_{^{12}CO(1-0)}$. Secondly, we note that 
Fig.~\ref{LCO2Lfir} shows a trend that the $L_{CO}$/$L_{FIR}$ ratio decreases with 
increasing dust temperature which is indicated by the two different FIR color ranges. 
This implies that the galaxies of the SLUGS sample which have higher luminosity is 
because they have a higher temperature in addition to simply more dust and gas, 
which would alone yield a linear relation between $L_{CO}$ and $L_{FIR}$. 
The correlation found between $L_{60 \mu m}$ and dust temperature by 
Dunne et al. (2000) also support this argument. This effect is consistent with 
the systematically higher $SFE$'s deduced for such objects, with the warmer 
IRAS colors, and with the saturation of the $r_{31}$ ratio at high 
$SFE$'s (see Fig.~\ref{r2sfe}).

As noted in \S~\ref{DnG}, $L_{FIR}$ is widely used as an indicator of the $SFR$, 
and $M$(H$_2$) is often computed from the CO flux. Thus, the non-linear relation 
between $L_{CO}$ and $L_{FIR}$ may reflect the behavior of star formation law 
for the nearby IR luminous galaxies. The star formation law can be represented as
a power law that can be expressed in terms of the star formation rate per unit area 
$\Sigma_{SFR}$ and the gas density $\Sigma_{gas}$, i.e. $\Sigma_{SFR}$ = 
$A$$\Sigma^N_{gas}$, first introduced by Schmidt (1959). Previous studies have 
shown that the activity of disk-averaged star formation is correlated with the mean
total molecular gas content \citep{ken89, bos94, bos95}, and this correlation is found 
to be stronger with H$_2$ gas in starburst regions \citep{krc98}. Fig.~\ref{sfr2mh2} 
shows a possible correlation between the $SFR$ per unit area and the H$_2$ gas density 
$\Sigma_{H_2}$ measured within a 15$^{\prime\prime}$ beam for our SLUGS sample. 
Both $\Sigma_{SFR}$ and $\Sigma_{H_2}$ are corrected for the inclination of galaxies 
in the SLUGS sample. The correction for the galaxy inclination is necessary because 
the Schmidt Law involves $\Sigma_{SFR}$ and $\Sigma_{H_2}$ when viewed normal to the 
disk. The disperson in the fit is about a factor of 3. The slope based on a linear 
fit in the log-log plane yields 1.4 $\pm$ 0.3. This value is consistent with the 
Schmidt law applied to the circumnuclear starbursts \citep{krc98}, though the 
uncertainty in the slope is too high to obtain a definitive value for the index. 
A possible reason for this is that our result refers to different regions in different 
galaxies, depending on the projected beamsize, whereas the work by Kennicutt and others 
consistently refer to disk-averaged parameters. The Schmidt Law involves H I+H$_2$, 
but ignoring H I gas, which averages about 30\% of the total gas mass measured within a 
15$^{\prime\prime}$ beam (see \S~\ref{gasratio}), is most likely not the cause of 
the scatter seen in Fig.~\ref{sfr2mh2}, since a study by Wong \& Blitz (2002) with 
a high angular resolution shows no correlation between H I and star formation rate. 

\subsection{Correlation between $r_{31}$ and Star Formation Parameters} \label{corrl}  

We examined in \S~\ref{r31} the excitation of CO and its relation with the properties 
of gas/dust and star formation in the central starburst regions in SLUGS sample. There 
are no significant correlations between $r_{31}$ and the distance of galaxies, star 
formation rate, dust temperature and mass, H$_2$ gas mass, the color indices, and 
the luminosities of IR and radio continuum. The lack of correlation of $r_{31}$ with 
properties relating to total star formation reflects a range of localized conditions 
in the molecular clouds. The possible dependence of $r_{31}$ on FIR/sub-mm surface 
brightness (see Fig.~\ref{r2sb}) and $SFE$ (see Fig.~\ref{r2sfe}) suggests that 
the higher degree of CO excitation is related to a higher concentration 
and efficiency of star forming activity. Such conditions would arise in an intense 
starburst where the surface density of such activity is high, consuming most of 
the gas in the region. The saturation effect of $r_{31}$ seen at higher $SFE$ which 
is probably due to thermalization of the CO levels, reflecting denser and warmer
gas in regions of high star formation efficiency. This interpretation is consistent 
also with the non-linear relation between $L_{CO}$ and $L_{FIR}$ 
(see Fig.~\ref{LCO2Lfir}), and with higher color temperatures associated with the 
most IR luminous and most efficient star forming galaxies. It is not clear whether 
these correlations reflect a high excitation ratio as the underlying cause or the 
effect of star forming activity. 

The regime where $r_{31}$ $>$ 1.0 requires optically thin CO lines from extremely 
warm and/or dense gas. The relatively small number of galaxies with high values 
of $r_{31}$ might be associated with the highly excited gas from a short-lived 
gas phase in the starburst systems. The lower line ratios may be associated 
with gas that is cooler and less denser. 

\subsection{The Molecular Gas Content} \label{gas}

The investigation of the gas and dust content depends critically on a reliable
estimate of the controversial parameter $X$. Studies have shown that this parameter 
varies from galaxy to galaxy \citep{bna98, bos02}, and it is thought to be higher
in metal-poor galaxies and lower in starburst galaxies than in Galactic molecular 
clouds, where $X$ is about 2.8 $\times$ 10$^{20}$ cm$^{-2}$ [K km s$^{-1}$]$^{-1}$
\citep{blo86, str88}. Thus in starburst galaxies, application of the standard factor 
can produce a significant overestimate of molecular hydrogen mass
\citep{sol97, dns98, zhu03}.

\subsubsection{The CO-to-H$_2$ Conversion Factor $X$ from An LVG Method} \label{X}

In this section, we examine the molecular gas content and properties using large 
velocity gradient (LVG) models. The LVG method is capable in principle of yielding 
the column density of CO, and hence of H$_2$, if an abundance ratio $Z_{CO}$ = 
[CO/H$_2$] is assumed. Hence, this method should yield an estimate of the $X$ factor 
independent of other considerations such as gas-to-dust ratio, which is discussed in 
the next section. We assume that the ISM comprises molecular clouds characterized by 
a single set of average physical properties. It is well understood that models 
incorporating more data for starburst galaxies require at least two components - 
a prevailing optically thin warm diffuse component, and an optically thick cool 
dense component with a smaller filling factor \citep[e.g.][]{aal95}. However, 
comparisons between single and double component models yield approximately the same 
column density for H$_2$, which is weighted strongly toward the prevailing diffuse 
component \citep[e.g.][]{zhu03}.

The LVG approach can be formulated into an equation yielding the $X$ factor from 
the LVG parameters given by 

\begin{equation}
X = \frac{n(H_2) \Lambda}{Z_{CO} T_{rad}} 
\end{equation}  

where  $n$(H$_2)$ is the H$_2$ gas density, $T_{rad}$ is the radiation temperature 
for the $^{12}$CO(1-0) line transition, $\Lambda$ = $\frac{Z_{CO}}{(dv/dr)}$, and 
$(dv/dr)$ is the velocity gradient associated with the individual clouds (see 
Appendix A). There are not enough data to uniquely determine the parameters in this 
equation, which usually require multiple line excitation ratios, measured brightness 
temperatures, and isotope intensity ratios to fully constrain the conditions in 
the molecular gas. In our case, it is therefore necessary to make some reasonable 
assumptions to constrain the LVG solutions. 

The highest degree of uncertainty concerns the parameter $Z_{CO}$ which is unknown
and needs to be addressed. Values widely quoted for starburst galaxies are in the 
range 10$^{-5}$ - 10$^{-4}$ \citep[e.g.][]{bna98, mao00}. Unfortunately, there are few 
if any reliable determinations for starburst galaxies. Within the Milky Way, 
measurements of $Z_{CO}$ and chemical models for dark and star forming clouds yield 
values within the range 5 $\times$ 10$^{-5}$ - 2.7 $\times$ 10$^{-4}$ 
\citep{bla87, far97, har98, van98, vnb98}. It seems unlikely therefore that 
$Z_{CO}$ $<$ 10$^{-5}$ for starburst galaxies unless their metallicity is unusually 
low, which would occur only in low luminosity systems. For the SLUGS sample, therefore, 
we adopt $Z_{CO}$ = 10$^{-4}$, bearing in mind the dependence of our computed values 
for $X$ on this parameter.

We adopt dust temperatures for each galaxy from Dunne et al. (2000), assuming 
$T_{dust}$ to be representative of gas kinetic temperature $T_{kin}$. 
Fig.~\ref{r2Td} shows a series of plots of $r_{31}$ versus $T_{dust}$ for our
subsample combined with theoretical curves based on LVG models. The theoretical 
curves represent $r_{31}$ versus $T_{kin}$ for a variety of assumed densities 
$n$(H$_2$) based on a series of LVG models using $Z_{CO}$ = 10$^{-4}$, 
and values of $\Lambda$ = 10$^{-7}$, 10$^{-6}$, 10$^{-5}$, and 10$^{-4}$ 
(km s$^{-1}$ pc$^{-1}$)$^{-1}$. In each case it is assumed that $T_{kin}$ = $T_{dust}$. 
Note the convergence of the curves at higher densities reflecting the saturation 
of the line ratios due to thermalization of the optically thick CO transitions. 
Note also that the ratios $r_{31}$ $>$ 1.0 can not be readily fit by the models with 
$\Lambda$ $\ge$ 10$^{-6}$ (km s$^{-1}$ pc$^{-1}$)$^{-1}$, a point to which we 
return later. 

Thus for each galaxy in one of the data plots, it is possible to derive a 
molecular gas density and radiation temperature $T_{rad}$ which is dependent
upon $\Lambda$. Using Equation (5), we then compute a range of $X$ parameter for 
the IR luminous galaxies. For each $\Lambda$, only points that fall below the upper 
limits of $r_{31}$ due to saturation are used for computing the distribution 
of $X$. For example, for $\Lambda$ = 10$^{-5}$ (km s$^{-1}$ pc$^{-1}$)$^{-1}$, 
thermalization occurs at $r_{31}$ = 0.86, corresponding to $T_{dust}$ = 55 K, 
$n$(H$_2$) = 2.4 $\times$ 10$^3$ cm$^{-3}$, and $T_{rad}$ = 27 K. The resulting 
range of $X$ is 1 - 3 $\times$ 10$^{19}$ cm$^{-2}$ [K km s$^{-1}$]$^{-1}$. 
For points falling above the saturation curves ($r_{31}$ $>$ 0.86) for $\Lambda$ = 
10$^{-5}$ (km s$^{-1}$ pc$^{-1}$)$^{-1}$ (see Fig.~\ref{r2Td}), we obtain $X$ 
values in a range 2.4 $\times$ 10$^{19}$ - 6.7 $\times$ 10$^{19}$ 
cm$^{-2}$ [K km s$^{-1}$]$^{-1}$ from the plots corresponding to 
$\Lambda$ = 10$^{-6}$ and 10$^{-7}$ (km s$^{-1}$ pc$^{-1}$)$^{-1}$. However, 
the latter value of $\Lambda$ is virtually unacceptable, since it corresponds 
to an impossibly high velocity gradient or impossibly low CO abundance. 
Therefore, $T_{kin}$ would have to be higher than the measured $T_{dust}$ for 
those points to lie below the saturation curves for $\Lambda$ $\le$ 10$^{-6}$ 
(km s$^{-1}$ pc$^{-1}$)$^{-1}$. Fig.~\ref{Xhis} shows the resulting distribution 
of the derived $X$ and its average values for different $\Lambda$. It is clear 
that regardless of $\Lambda$, the values of $X$ derived from starburst galaxies 
are systematically lower than the value $X$ = 2.8 $\times$ 10$^{20}$ 
cm$^{-2}$ [K km s$^{-1}$]$^{-1}$ derived from the disk of the Galaxy.

In order to further constrain the range of the $X$ factor and $\Lambda$ shown in 
Fig.~\ref{Xhis}, we compare the CO isotope intensity ratios 
$R_{10}$  = $I$[$^{12}$CO(1-0)]/$I$[$^{13}$CO(1-0)] predicted by the LVG model 
with observed values. We adopted the isotope abundance ratio as 
$\Big[\frac{^{12}CO}{^{13}CO}\Big]$ = 40 - 75 \citep{wnr94}. The CO isotope 
intensity ratios data are taken from Aalto et al. (1995) and Taniguchi et al. (1999) 
for 34 IR luminous galaxies which have a range of FIR luminosities similar to the 
SLUGS sample, i.e. $L_{FIR}$ $>$ 10$^{10}$ L$\odot$. Fig.~\ref{R10his} shows a 
comparison between the distributions of computed and observed isotope ratios for the 
sample as a whole for each value of $\Lambda$, and corresponding value for the mean 
derived $X$ factor. The plots show that the best agreement with the observed 
distribution of $R_{10}$ is obtained for $\Lambda$ = 10$^{-5}$ 
(km s$^{-1}$ pc$^{-1}$)$^{-1}$, corresponding to an average value for $X$ = 2.7 
$\times$ 10$^{19}$ cm$^{-2}$ [K km s$^{-1}$]$^{-1}$. The other plots assuming 
other values of $\Lambda$ show marked disagreement with the observed isotope line 
ratios. 

The resulting value for $X$ (2.7 $\times$ 10$^{19}$ cm$^{-2}$ [K km s$^{-1}$]$^{-1}$) 
in starburst regions is thus ten times smaller than the conventional value of $X$ 
for GMC's in the Milky Way. But the result is comparable with $X$ = 0.5 $\times$
10$^{20}$ cm$^{-2}$ [K km s$^{-1}$]$^{-1}$ estimated from diffuse clouds in the 
Galaxy by Polk et al. (1988), and with that found for extreme starbursts in nearby 
galaxies by Downes \& Solomon (1998) and for starburst galaxy M 82 by 
Mao et al. (2000).  

We repeated the LVG model calculations for a higher CO isotope abundance ratio of 75 
\citep{wnr94}, and found that the average value of $X$ for $Z_{CO}$ = 10$^{-4}$ 
becomes 8.2 $\times$ 10$^{19}$ cm$^{-2}$ [K km s$^{-1}$]$^{-1}$. We also repeated 
the entire analysis by increasing the adopted kinetic temperatures by 50\%, and found 
that the resulting value of $X$ remains essentially the same, since for a given 
$\Lambda$, the lower densities required are largely offset by lower values of the 
radiation temperature (see Equation (5)). Our LVG analysis also shows that the 
predicted isotope ratios $^{12}$CO(2-1)/$^{13}$CO(2-1) are in satisfactory agreement 
with the observed ratios reported by Taniguchi et al. (1999). The results do not 
seem to support the conclusion by Taniguchi et al. (1999) that the comparison 
between CO(2-1) and CO(1-0) isotope intensity ratios may imply that the unusually 
large values in starburst systems indicate an isotope abundance anomaly in starburst 
galaxies. Our results are consistent with effects of excitation and optical depth.  

\subsubsection{The $M$(H$_2$)/$M$(H I) Ratio} \label{gasratio}

In this section, we briefly consider a small subset of our sample with H I maps 
available to assess the combined molecular and atomic gas content within the telescope 
beam. The subset comprises twelve galaxies with an average 
$L_{FIR}$ = 4.1 $\times$ 10$^{10}$ L$_{\odot}$ \citep{tho02}. The H I masses for our 
15$^{\prime\prime}$ beam were estimated from the peak H I column densities maps at 
25$^{\prime\prime}$ resolution by assuming a uniform H I brightness distribution within 
the 25$^{\prime\prime}$ beam. The mean ($\pm$ r.m.s.) projected beamsize for the subset 
is 3.6 ($\pm$ 1.1) kpc.

The overall mean ($\pm$ r.m.s.) values for $M$(H$_2$)/$M$(H I) ratio and combined 
H$_2$+H I face-on surface density are respectively 2.0 $\pm$ 1.5 and 42 $\pm$ 18 
M$_{\odot}$ pc$^{-2}$.  These quantities must be considered respectively as upper and 
lower limits however, since the H I masses may be underestimated due to the effects
of absorption by H I against the radio continuum in the nuclear regions \citep{tho02}. 
This is particularly the case for one galaxy, NGC 5900, for which $M$(H$_2$)/$M$(H I) 
= 21, and this object was omitted from the mean value quoted above. The mean ratio 
$M$(H$_2$)/$M$(H I) is substantially larger than that found for optically selected 
late type galaxies (0.15) by Boselli et al (2002). This is partly attributable to 
the fact that their values are global measures, whereas our values refer to the 
nuclear regions of galaxies selected according to high star formation rates. 
The H$_2$+H I surface density is intermediate between the central surface densities 
for galaxies with lower star forming rates like the Milky Way, which is about 
10 M$_{\odot}$ pc$^{-2}$ \citep{elm89} and those for ultra-luminous IR galaxies, 
which are up to 10$^5$ M $_{\odot}$ pc$^{-2}$ \citep{aal95}.

It should be emphasized that these estimates refer to comparatively warm gas 
heated by star formation in these systems. The outer disks of galaxies may be hiding 
substantial amounts of cool H$_2$, where the radiative efficiency of CO is low (i.e. 
$X$ is high). The $M$(H$_2$)/$M_{dust}$ ratio will be discussed in a separate paper
(in preparation), where the dust mass of the SLUGS sample \citep{dun00} will be 
corrected for the effects of contamination of the SUCBA 850 $\mu$m filter.

\subsubsection{Virial Stability of The Molecular Clouds} \label{vir}

Using the virial theorem, the velocity gradient under the virial condition for a 
spherical cloud with a mean gas density $<n>$ is given by, 

\begin{equation}
\Big(\frac{dv}{dr}\Big)_{VIR} = 0.65 \alpha^{\frac{1}{2}} \Big(\frac{<n>}{10^3 cm^{-3}}\Big)^{\frac{1}{2}} km s^{-1} pc^{-1}
\end{equation}

where $\alpha$ is in the range 0.5 - 2.5 \citep{pap98, bns96} depending primarily 
on the assumed density profile. For a mean gas density of 1.4 $\times$ 10$^3$ cm$^{-3}$
for $r_{31}$ $\le$ 0.8 derived from the LVG model for our IR luminous galaxies (see 
Fig.~\ref{r2Td}) with $\Lambda$ = 10$^{-5}$ (km s$^{-1}$ pc$^{-1}$)$^{-1}$ and 
$Z_{CO}$ = 10$^{-4}$, we estimate a virial velocity gradient of 
1.1 km s$^{-1}$ pc$^{-1}$ for $\alpha$ = 2.5. This virial value is about 9 - 90 times 
smaller than the velocity gradient of 10 - 100 km s$^{-1}$ pc$^{-1}$ also derived from 
LVG modeling using $\Lambda$ = 10$^{-5}$ - 10$^{-6}$ (km s$^{-1}$ pc$^{-1}$)$^{-1}$ 
and $Z_{CO}$ = 10$^{-4}$. The same calculation was made for a lower $r_{31}$ using
the source UGC 6436 with $r_{31}$ = 0.35, and a corresponding gas density of 
7.5 $\times$ 10$^2$ cm$^{-3}$. It was found that $\Big(\frac{dv}{dr}\Big)_{VIR}$ is 
still about 11 times smaller than the LVG velocity gradient. Since such low values of 
$r_{31}$ are more sensitive to density than more saturated values near unity, this 
result provides a strong confirmation of the non-virial expansion velocities. 
As noted earlier, there are a number of galaxies with high $^{12}$CO line 
ratios ($r_{31}$ $\sim$ 0.8 - 1.72) seen in Fig.~\ref{r2Td}, where $\Lambda$ may 
be lower than 10$^{-5}$ (km s$^{-1}$ pc$^{-1}$)$^{-1}$. The above estimates imply 
that the molecular clouds in the starburst regions are not gravitationally bound, 
confirming the suggestion recently made by Zhu et al. (2003, ApJ in press) for 
the Antennae galaxies, unless one is willing to accept a significantly lower value 
for $Z_{CO}$. The cause could be the destruction by the stellar winds and expanding 
shells of new-born massive stars. We note that the non-virialized component emanates 
from the gas phase that dominates the $^{12}$CO emission, and not necessarily from 
the entire cloud mass, in accordance with two-phase gas models \citep{aal95, pns99}.  

\section{Conclusions}

The main conclusions of this work are:

1. The $L_{CO}$/$L_{FIR}$ ratios measured from the SLUGS sample decrease with 
increasing FIR luminosity. The non-linearity between $L_{CO}$/$L_{FIR}$ and $L_{FIR}$
implies that the more IR luminous galaxies have higher dust temperatures than those
at the low luminosity end. This non-linear relation was found to be consistent
with that expect for the Schmidt Law for star formation with an index 
$N$ = 1.4 $\pm$ 0.3, though the index is too uncertain to provide a new and 
definitive Schmidt relation.

2. The degree of CO excitation measured by the $^{12}$CO(3-2)/(1-0) line intensity 
ratios vary greatly from galaxy to galaxy in the SLUGS sample. There is a trend for 
the $r_{31}$ ratios to increase with increasing concentration and efficiency of star 
forming activity. The saturation of $r_{31}$ seen at higher $SFE$ implies that 
the gas is denser and warmer in regions of high $SFE$ which is consistent with the 
non-linearity between $L_{CO}$/$L_{FIR}$ and $L_{FIR}$. 

3. Our $^{12}$CO line measurements together with dust and CO isotope data taken from 
the literature are modeled using the LVG method to estimate that the CO-to-H$_2$ 
conversion factor lies between 2.7 $\times$ 10$^{19}$ and 6.6 $\times$ 10$^{19}$ 
cm$^{-2}$[K km s$^{-1}$]$^{-1}$ for SLUGS galaxies, which is about 4 - 10 times lower 
than the conventional $X$ derived from the Galaxy. 

4. For a subset of 12 galaxies with available H I maps, the total gas face-on surface 
density within about 2 kpc of the center averages about 42 M$_{\odot}$ pc$^{-2}$, which 
is intermediate between that for disk galaxies with low and high star formation rates.

5. Our Virial analysis shows that the molecular clouds in starburst regions are not 
gravitationally bound unless one is willing to accept a 9 - 90 times lower [CO/H$_2$] 
abundance ratio. Most of the $^{12}$CO line emission originates from the non-virialized
warm and diffuse gas clouds.  

\acknowledgments

We thank the staffs of the JCMT and NRO, and Dr. Siow-wang Lee at University of 
Toronto for their assistance with the observations. We thank Alan Richardson
at University of Toronto for his assistance in the analysis of SCUBA 850 $\mu$m and 
cm radio images. We also thank Dr. Helen Thomas at University of Cambridge 
for providing us the electronic forms of twelve H I maps (from Thomas et al. 2002). 
L.Y. would like to thank Dr. Phil Arras at Kavli Institute for Theoretical Physics. 
We thank the referee for many helpful comments. This research was supported by a 
research grant from the Natural Sciences and Engineering Research Council of Canada, 
and a Reinhardt Graduate Student Travel Fellowship from the Department of Astronomy 
and Astrophysics at the University of Toronto.

\appendix

\section{Derivation of The CO-to-H$_2$ Conversion Factor $X$ From LVG Model}

From the LVG model for a spherical molecular cloud, the column density per unit 
velocity range in the average cloud is given by

\begin{equation}
n(H_2) \Lambda = \frac{n(H_2) Z_{CO}}{dv/dr} 
\end{equation}

where $n$(H$_2$) is the density of H$_2$, $\Lambda$ = $\frac{Z_{CO}}{dv/dr}$, 
$Z_{CO}$ is the CO abundance ratio [CO/H$_2$], and $dv/dr$ is the velocity 
gradient within the cloud. The total CO column density per cloud $N_{CO}$, 
integrated over velocity is then

\begin{equation}
N_{CO} = n(H_2) \Lambda \Delta v
\end{equation}

where $\Delta v$ is the velocity width of an individual cloud. 

The column density $\overline{N_{CO}}$ averaged over the beam area can be written as 

\begin{equation}
\overline{N_{CO}} = n(H_2) \Lambda \Delta v \frac{N_{cloud} \sigma}{A_{beam}}
\end{equation}

or, 

\begin{equation}
\overline{N_{CO}} = n(H_2) \Lambda \Big[\frac{N_{cloud} \sigma}{A_{beam}} \frac{\Delta v}{\Delta V}\Big] \Delta V
\end{equation}

where $N_{cloud}$ is the number of clouds in the beam, $\sigma$ is the cloud cross 
section area, $A_{beam}$ is the beam area, and $\Delta V$ is the line width of all 
emission detected in the beam. 

The quantity inside inside the brackets of Equation (A4) is the brightness dilution 
factor representing the product of the geometric and velocity filling factors within 
the beam.

Assuming for simplicity, a rectangular line profile, then

\begin{equation}
\overline{N_{CO}} = n(H_2) \Lambda \Big[\frac{T_{mb}}{T_{rad}}\Big] \Delta V
\end{equation}

where the radiation temperature $T_{rad}$ is calculated from LVG modeling for
$^{12}$CO(1-0) or (3-2) line emission, and $T_{mb}$ is the corresponding
main-beam temperature.

The quantity $T_{mb} \Delta V$ can be replaced by $\int T_{mb} dV$, the line flux in
units of brightness. Thus

\begin{equation}
\overline{N_{CO}} = \frac{n(H_2) \Lambda}{T_{rad}} \int T_{mb} dV
\end{equation}

Setting $\overline{N_{CO}}$ = $Z_{CO}$ $\overline{N(H_2)}$, where 
$\overline{N(H_2)}$ is the corresponding column density of H$_2$, and 
$X$ = $\frac{\overline{N(H_2)}}{I_{10}}$, where 

\begin{equation}
I_{10} =  \int T_{mb} dV 
\end{equation}

for $^{12}$CO (1-0) line emission. We obtain

\begin{equation}
X = \frac{n(H_2) \Lambda}{Z_{CO} T_{rad}}.
\end{equation}

\clearpage

\setcounter{figure}{0}
\begin{figure}
\plotone{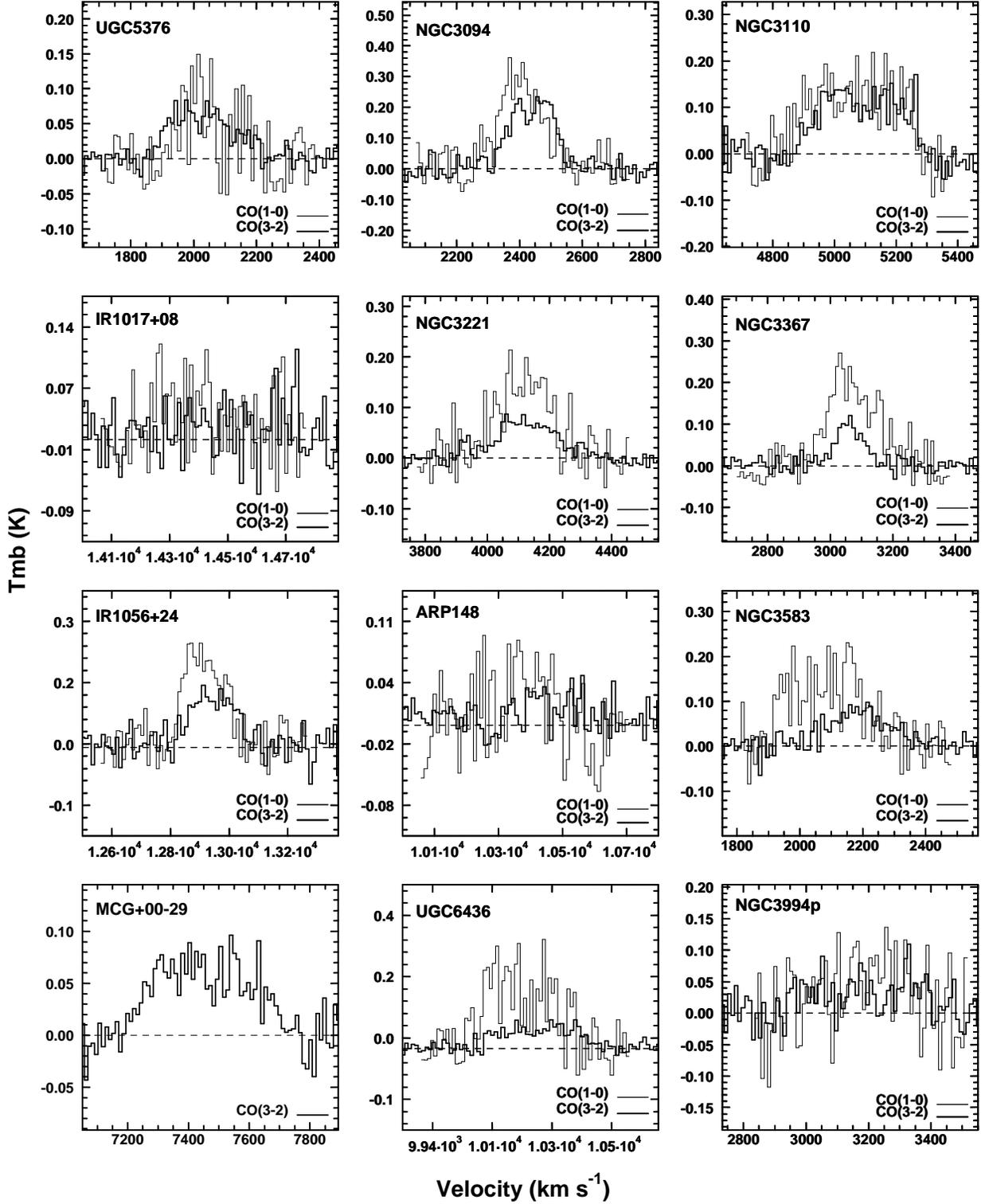}
\caption{Spectra of $^{12}$CO(1-0) and $^{12}$CO(3-2) for individual sources.
The velocity scale is Heliocentric. The temperature scale is in units of main beam 
temperature. The letter p denotes those galaxies that belong to a pair. 
\label{cospectra}}
\end{figure}

\setcounter{figure}{0}
\begin{figure}
\plotone{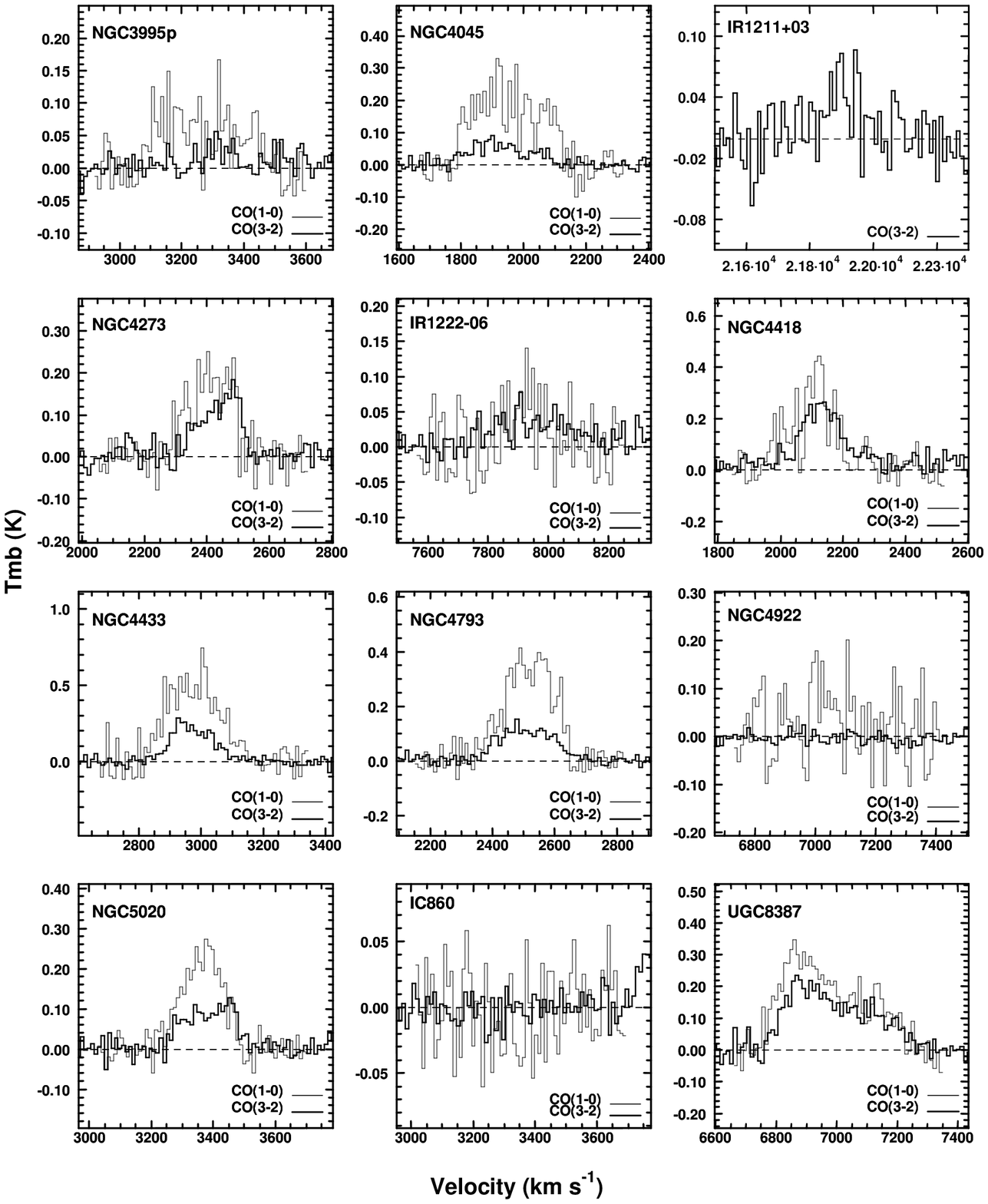}
\caption{(continued)
}
\end{figure}

\setcounter{figure}{0}
\begin{figure}
\plotone{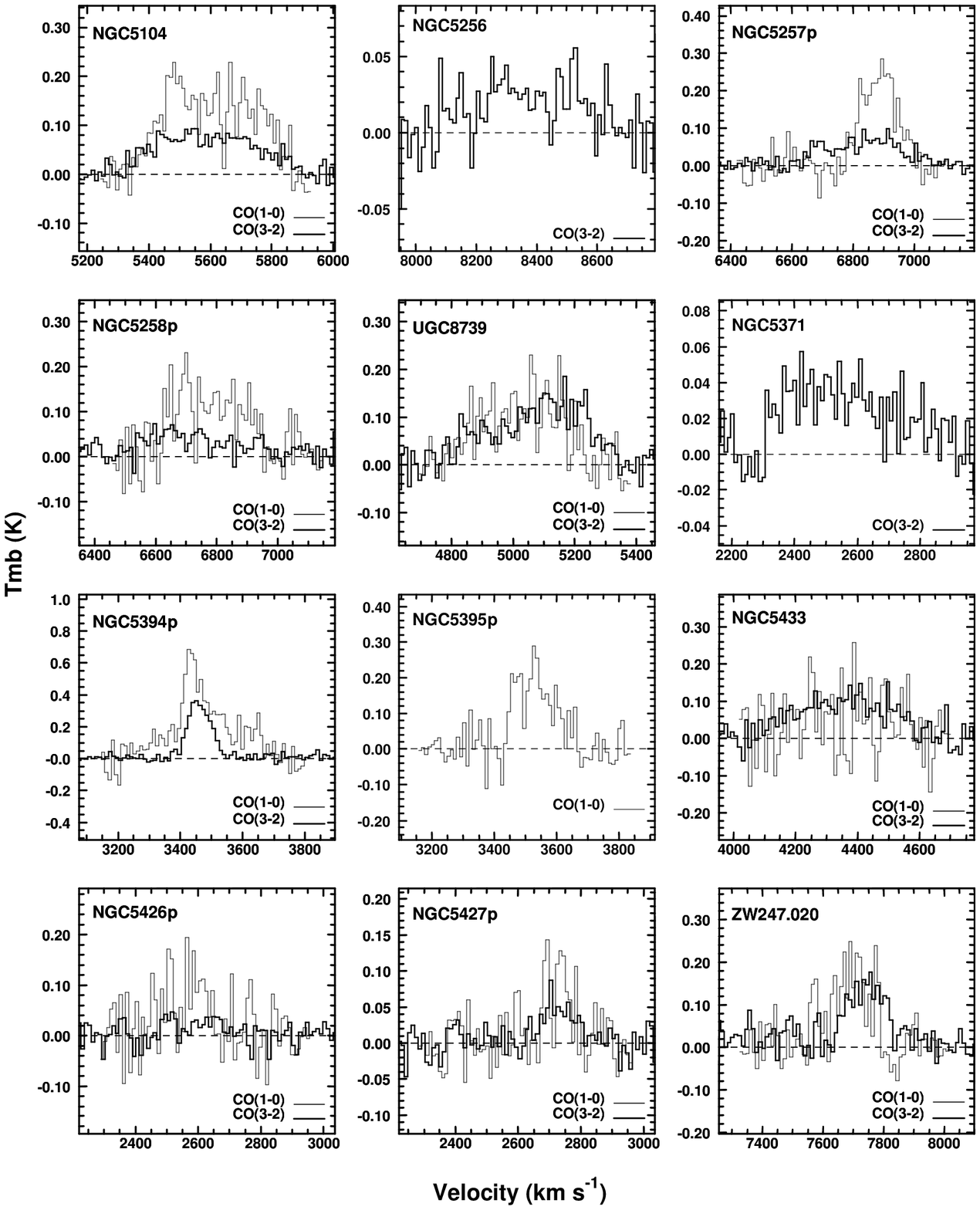}
\caption{(continued) 
}
\end{figure}

\setcounter{figure}{0}
\begin{figure}
\plotone{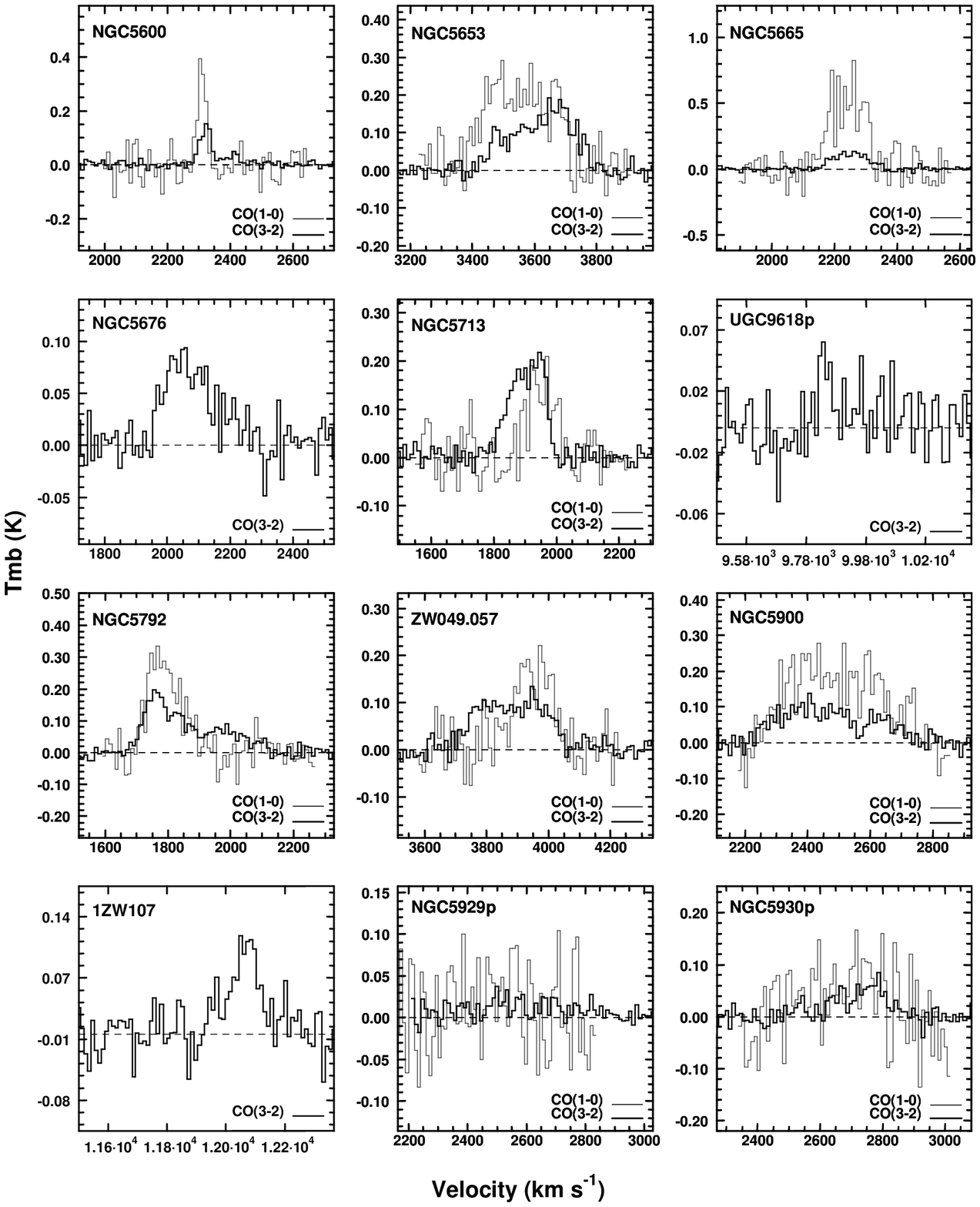}
\caption{(continued) 
}
\end{figure}

\setcounter{figure}{0}
\begin{figure}
\plotone{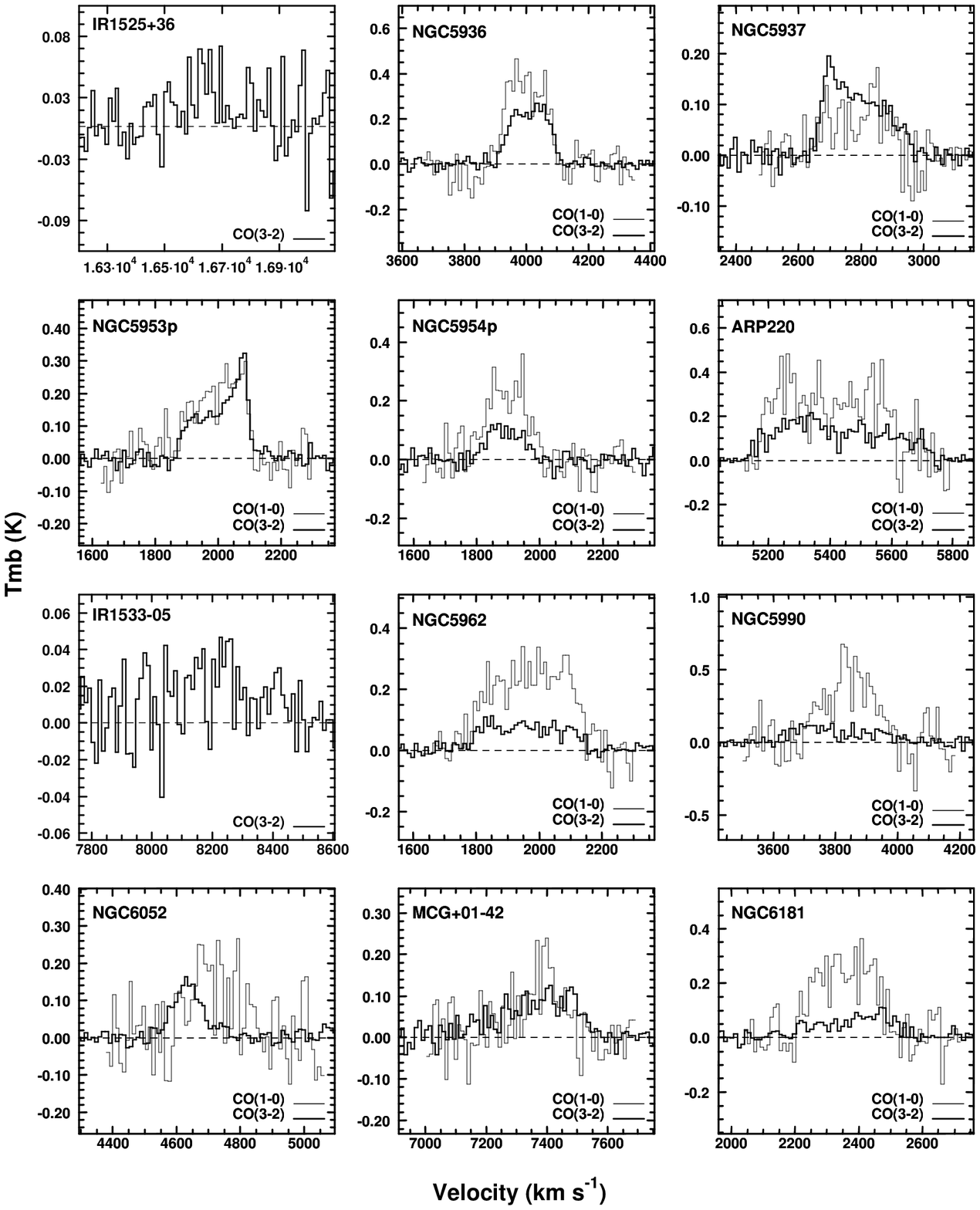}
\caption{(continued) 
}
\end{figure}

\clearpage

\begin{figure}
\epsscale{.60}
\plotone{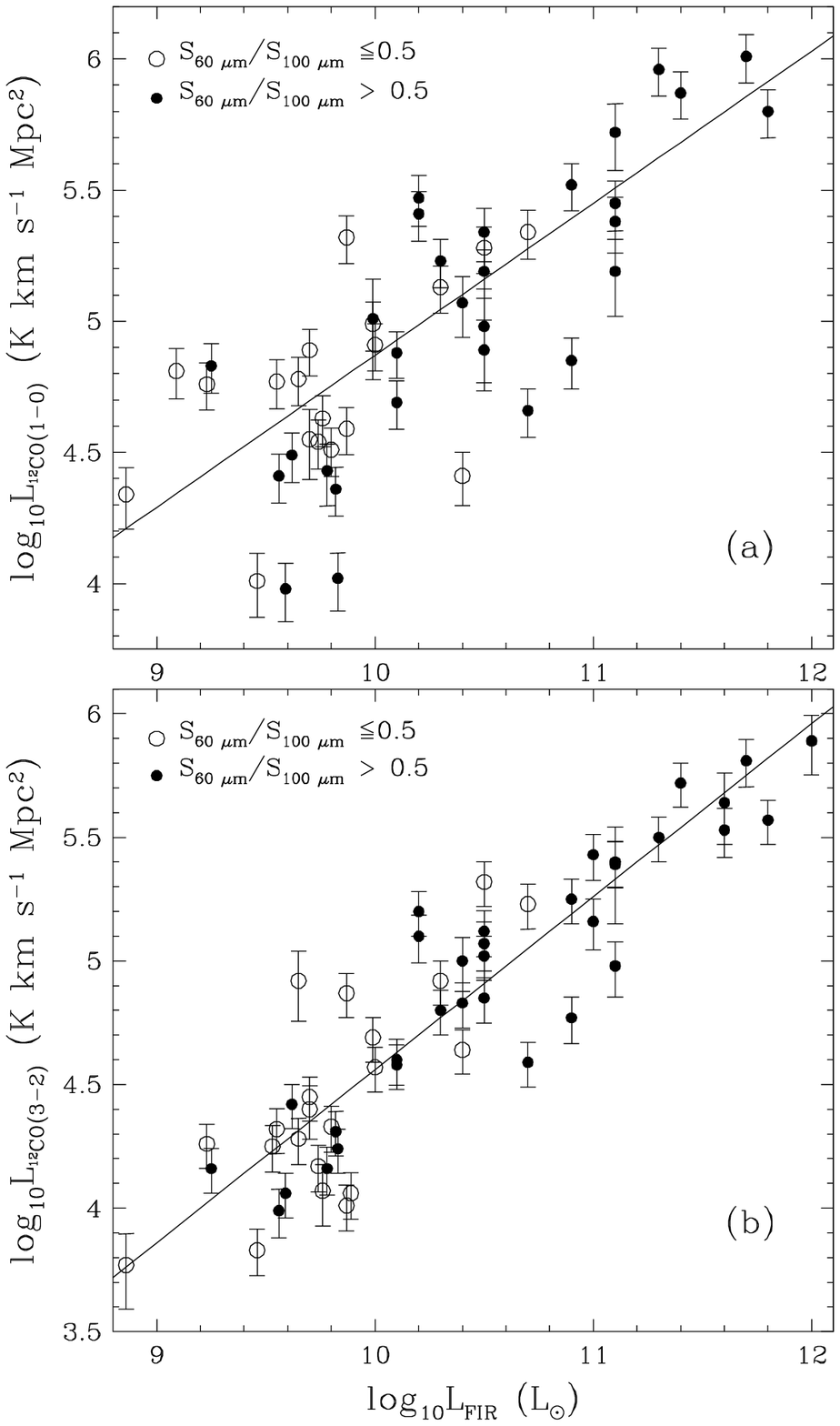}
\caption{Plots of CO luminosity versus FIR luminosity, both associated with the
15$^{\prime\prime}$ beam for (a) $^{12}$CO(1-0) and (b) $^{12}$CO(3-2) emission. 
The lines represent linear regression fits to the data. The data points have been 
segregated according to two FIR color regimes in the plot. See text for details.
\label{LCO2Lfir}}
\end{figure}

\clearpage

\begin{figure}
\epsscale{.60}
\plotone{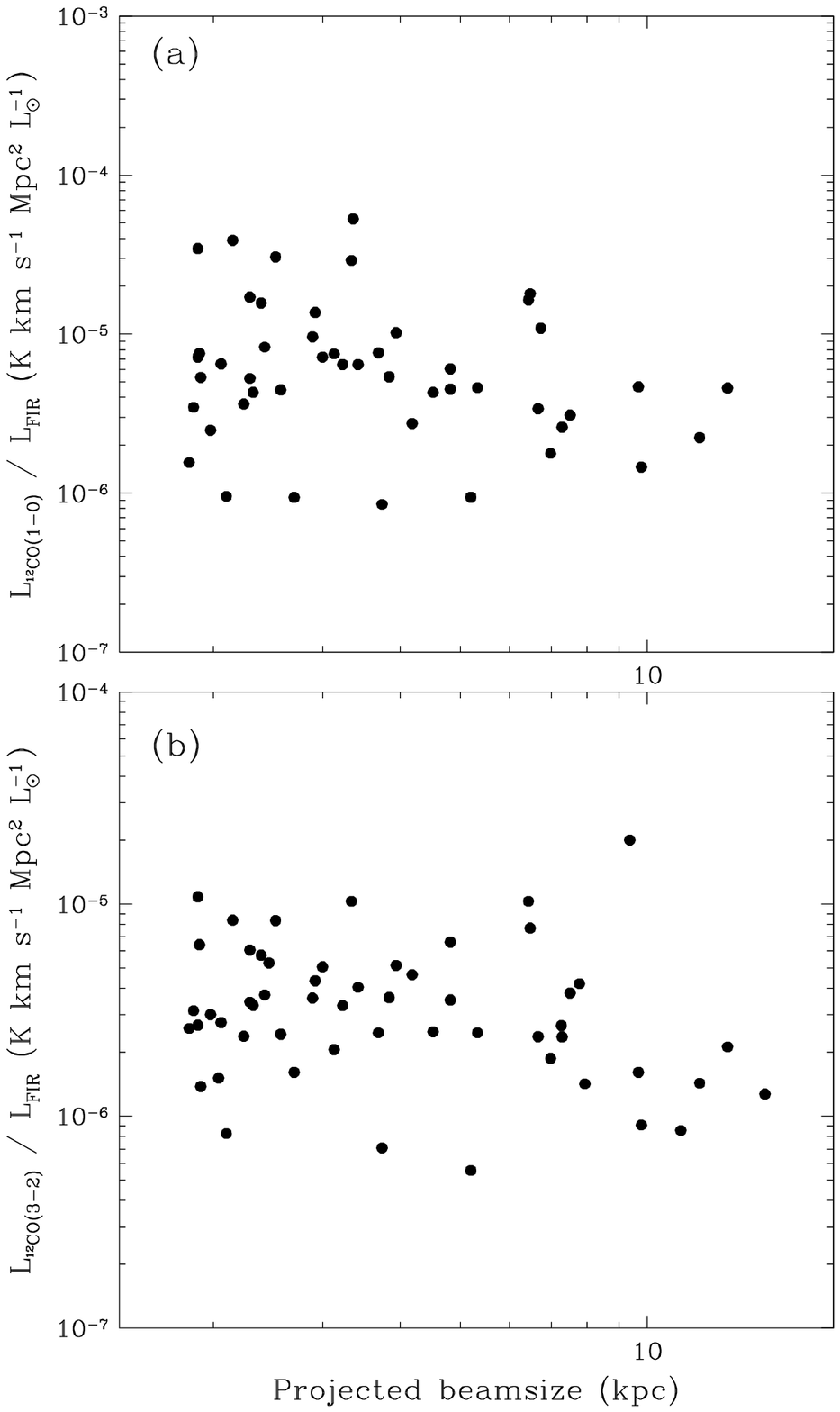}
\caption{Plots of the ratio of CO-to-FIR luminosity versus the projected beamsize 
for (a) $^{12}$CO(1-0) and (b) $^{12}$CO(3-2) emission.  
\label{rL2l}}
\end{figure}

\clearpage

\begin{figure}
\plotone{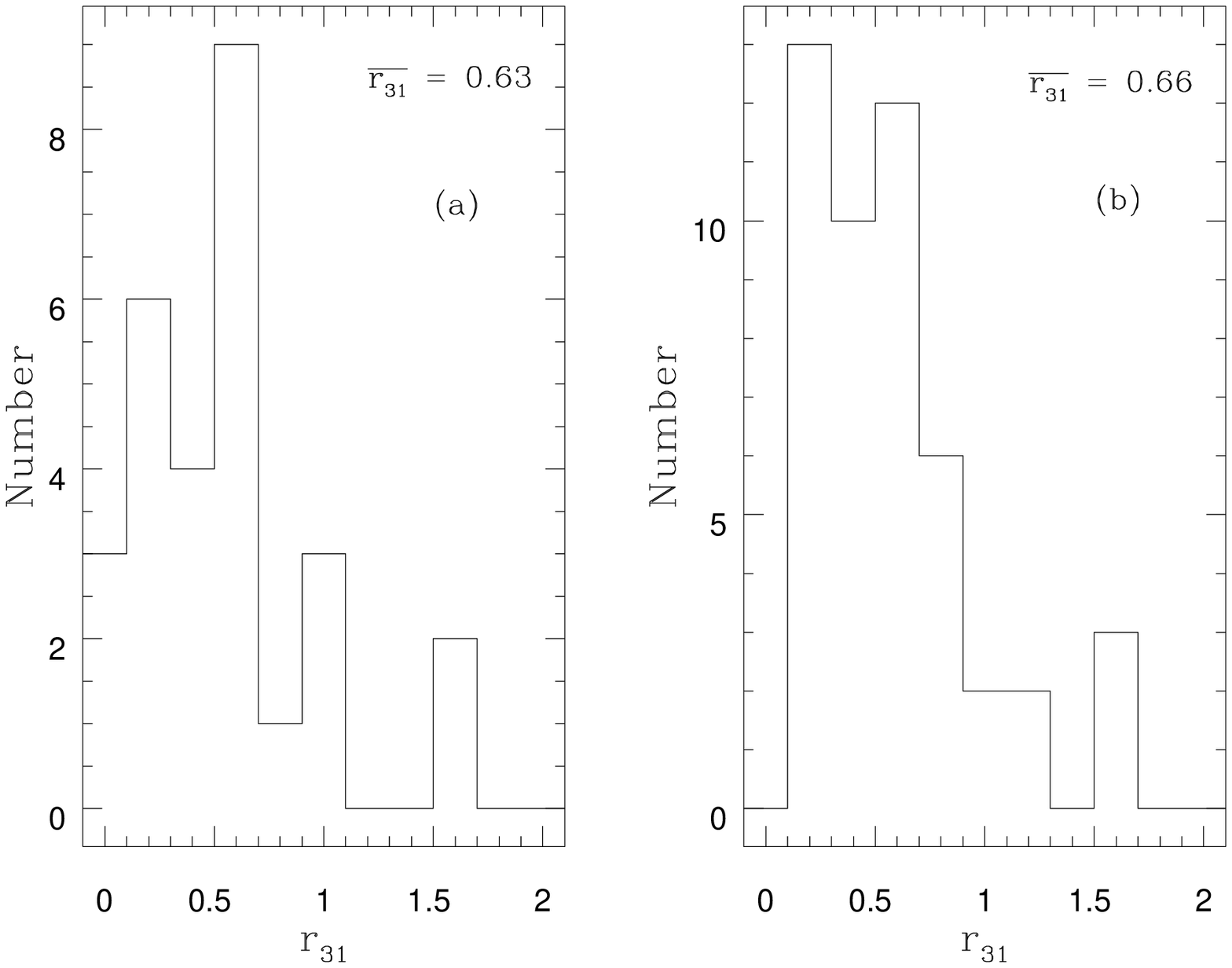}
\caption{Distribution of the ratio $r_{31}$ for IR luminous galaxies, (a) by 
Mauersberger et al. (1999); (b) this paper. 
\label{r31his}}
\end{figure}

\clearpage

\begin{figure}
\plotone{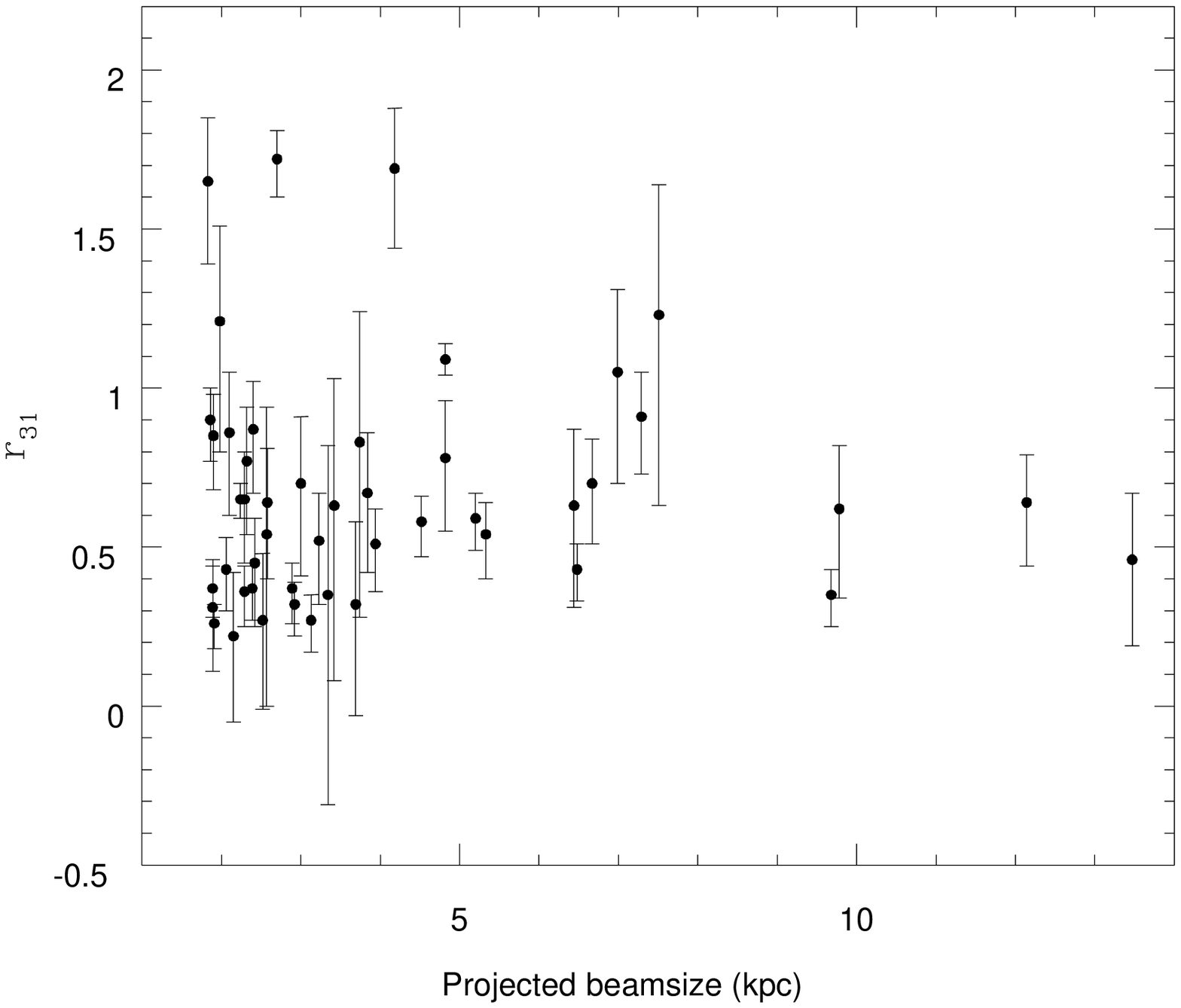}
\caption{Plot of $r_{31}$ versus the projected beamsize for the SLUGS subsample. 
\label{r2l}}
\end{figure}

\clearpage 

\begin{figure}
\epsscale{.60}
\plotone{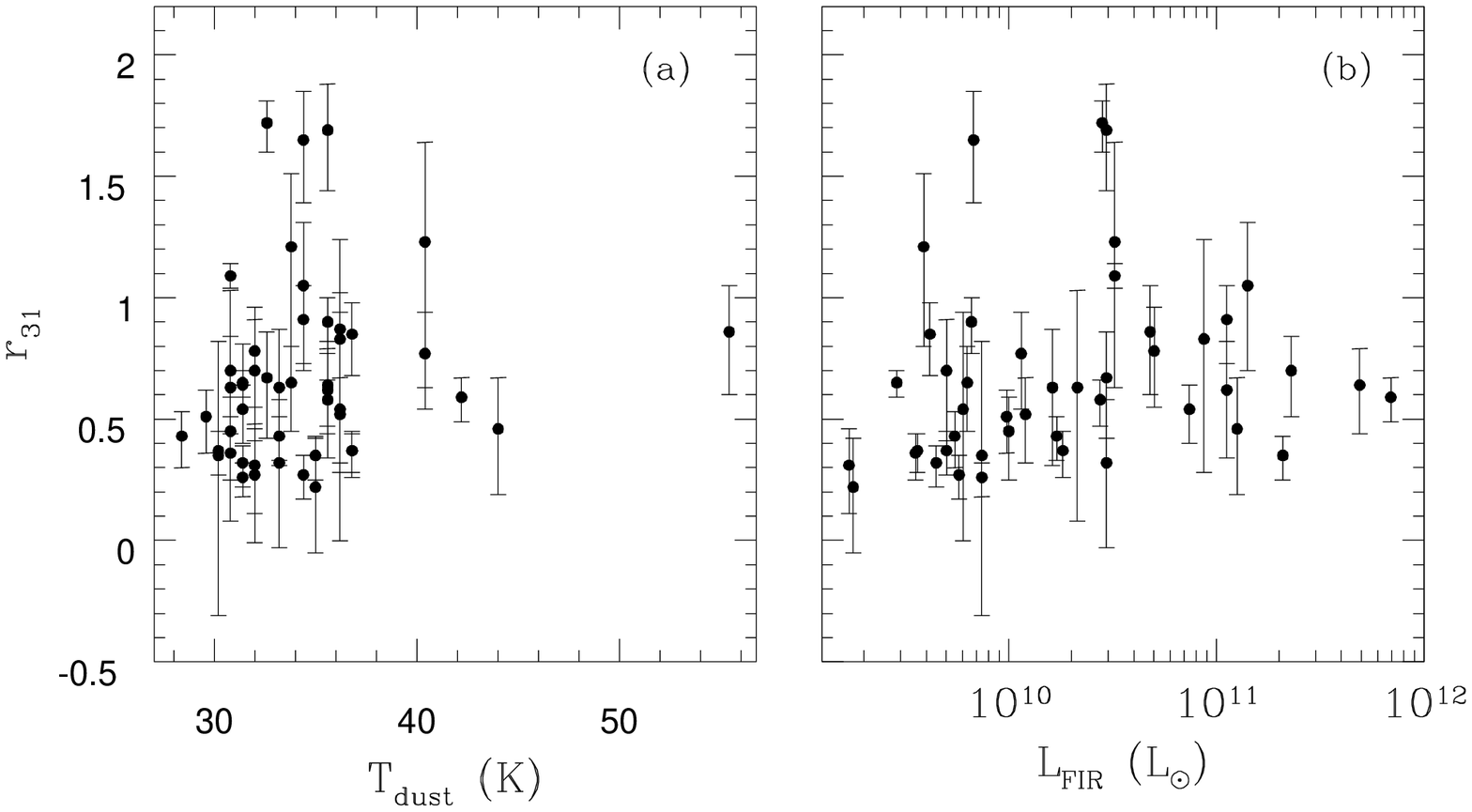}
\caption{Plots of (a) $r_{31}$ versus the dust temperature $T_{dust}$, (b) $r_{31}$ 
versus the FIR luminosity $L_{FIR}$ applicable to the 15$^{\prime\prime}$ beam. 
\label{r2TdLfir}}
\end{figure}

\clearpage 

\begin{figure}
\plotone{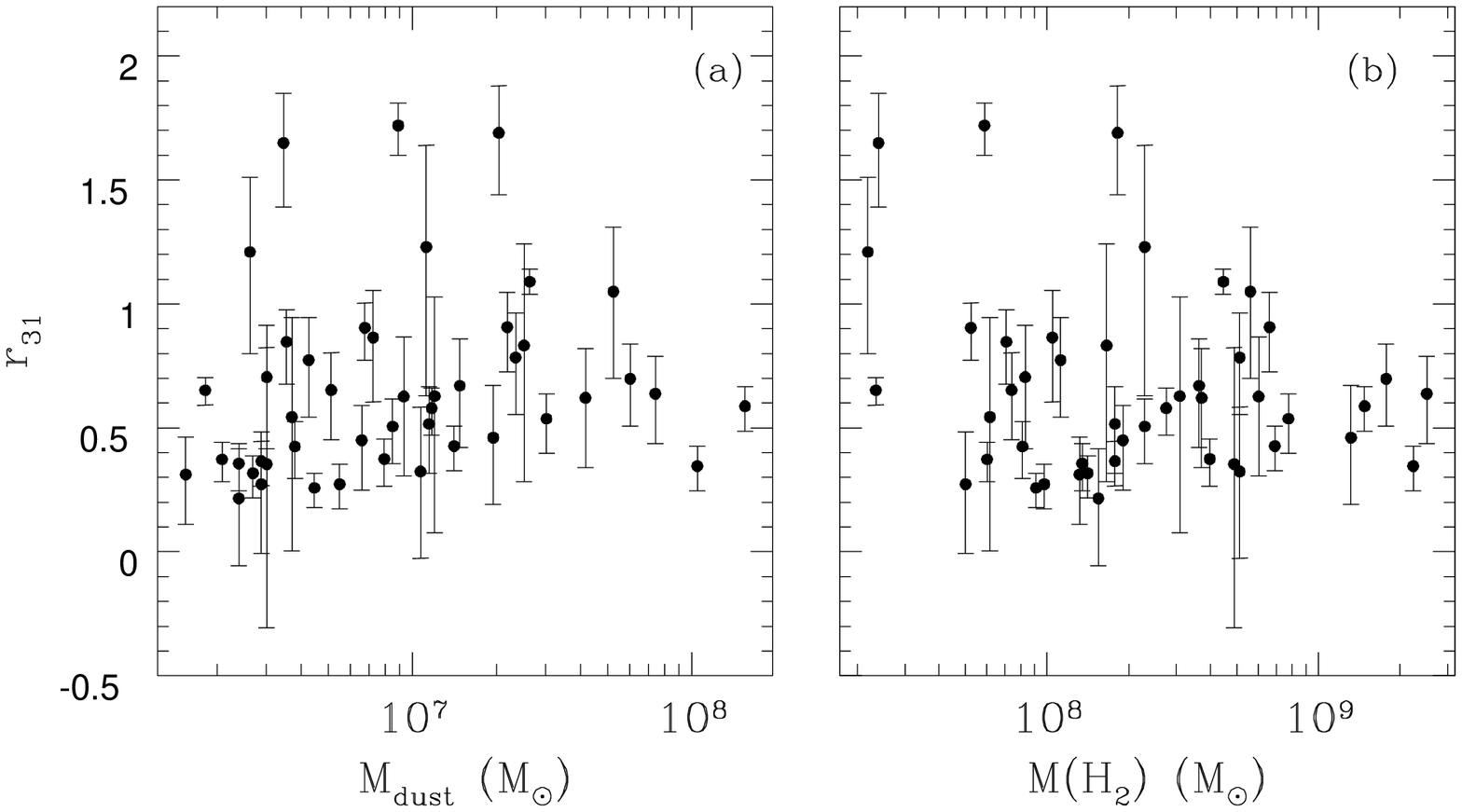}
\caption{Plots of (a) $r_{31}$ versus dust mass $M_{dust}$, (b) $r_{31}$ versus
molecular gas mass $M$(H$_2$) applicable to the 15$^{\prime\prime}$ beam. 
\label{r2M}}
\end{figure}

\clearpage

\begin{figure}
\epsscale{.60}
\plotone{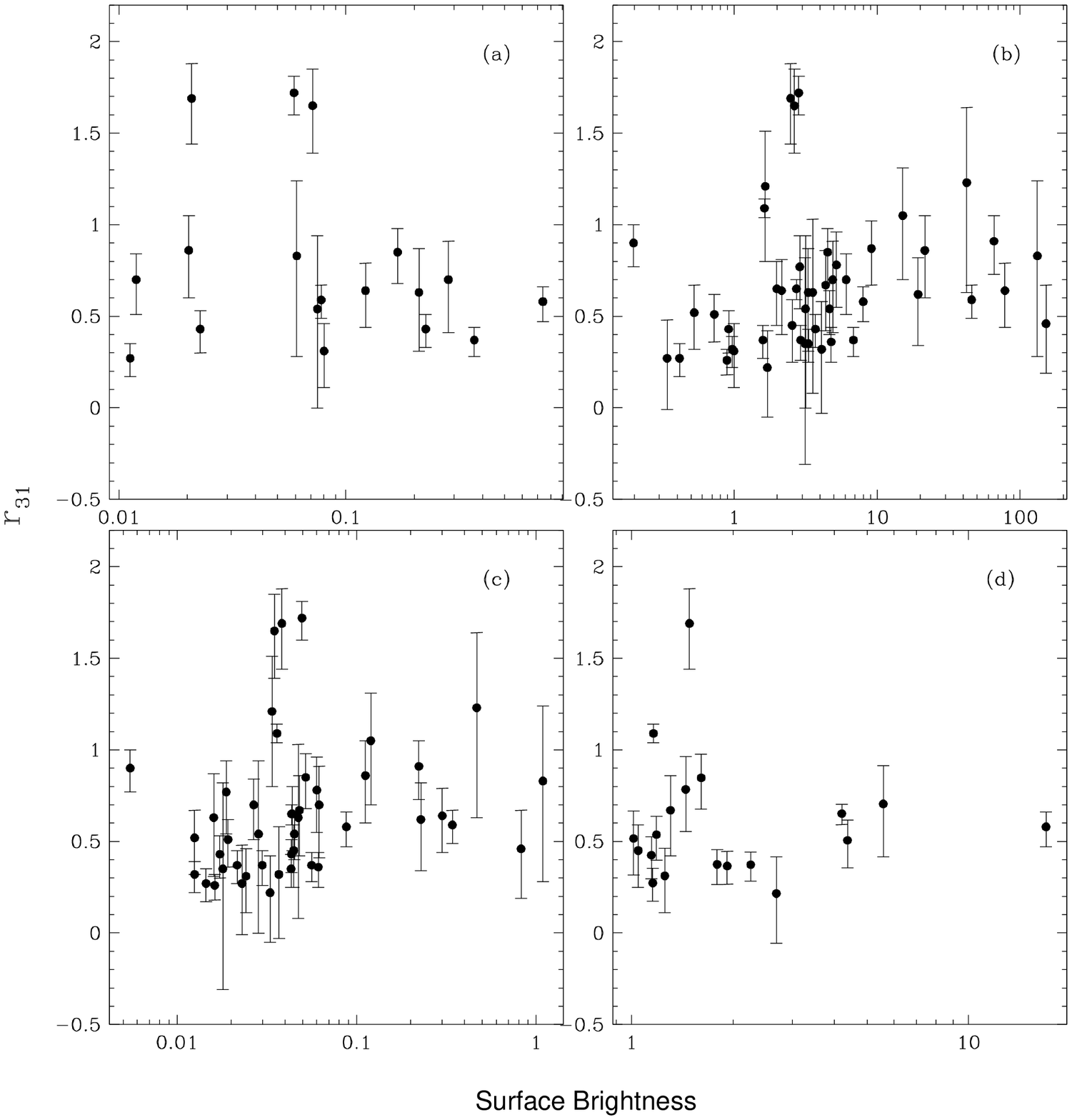}
\caption{Plots showing $r_{31}$ versus the surface brightness averaged over the optical 
diameter for (a) mid-IR 14.3 $\mu$m, (b) far-IR 60 $\mu$m, (c) SCUBA 850 $\mu$m, and 
(d) H I 21cm line. 
\label{r2sb}}
\end{figure}

\clearpage 

\begin{figure}
\epsscale{.60}
\plotone{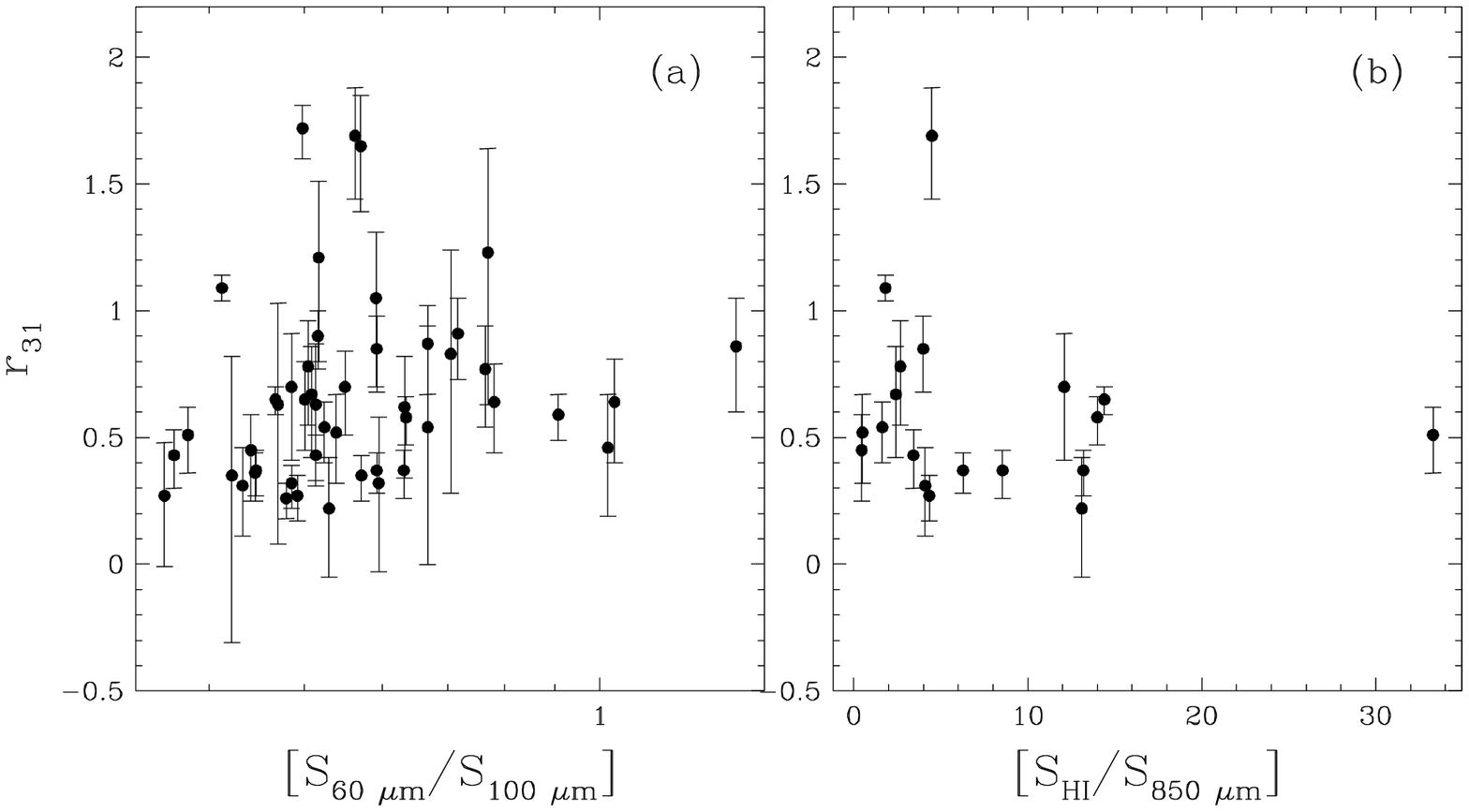}
\caption{Plots of $r_{31}$ versus the color indices or ratios involving (a) far-IR and 
(b) sub-mm to H I 21cm emission. 
\label{r2cidx}}
\end{figure}

\clearpage

\begin{figure}
\plotone{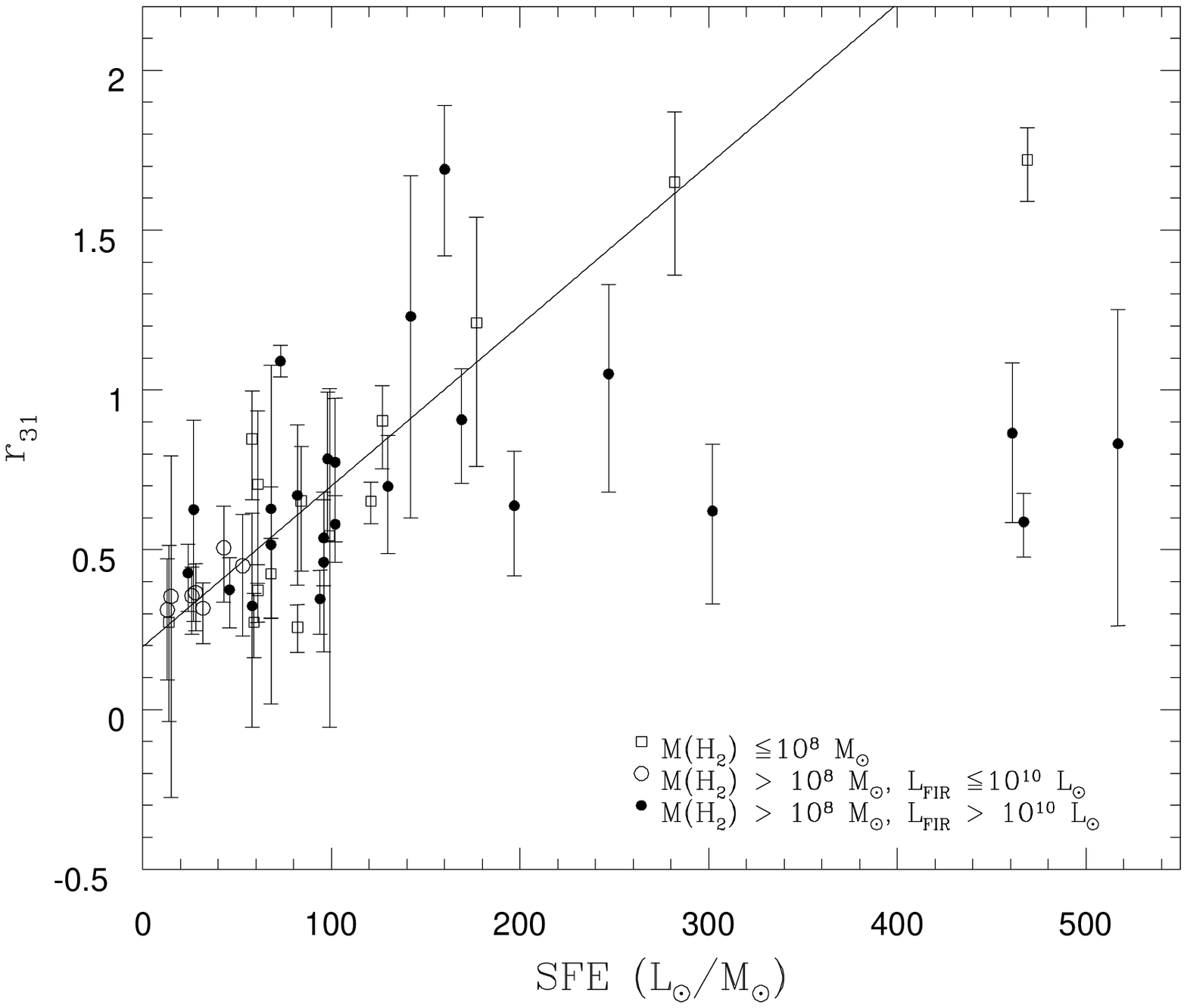}
\caption{Plot of $r_{31}$ versus the star formation efficiency $SFE$ = 
$L_{FIR}$/$M$(H$_2$) measured within a 15$^{\prime\prime}$ beam. The line represents
a linear regression fit to the data with $SFE$ $\le$ 200 L$_{\odot}$/M$_{\odot}$. 
The data points have been segregated according to gas mass-class and dust IR 
luminosity-class to show the relationship between $L_{FIR}$ and $M$(H$_2$) and position 
in the plot. See text for details.
\label{r2sfe}}
\end{figure}

\clearpage

\begin{figure}
\plotone{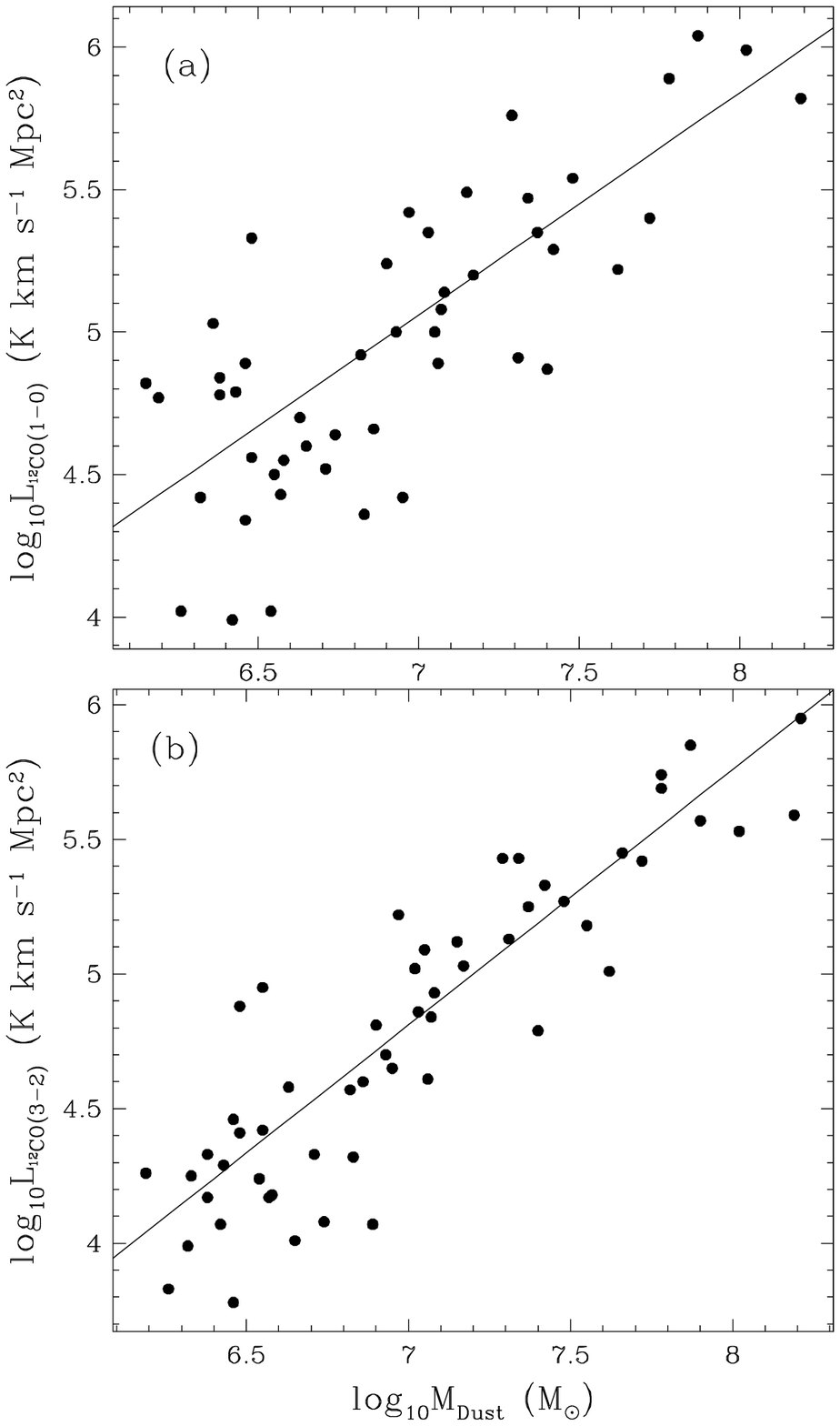}
\caption{Plots of CO luminosities for the 15$^{\prime\prime}$ beam versus dust mass 
from Dunne et al. (2000) for (a) $^{12}$CO(1-0) and (b) $^{12}$CO(3-2) 
emission. The dust masses have not been corrected for contamination of the SCUBA 
850 $\mu$m filter by $^{12}$CO(3-2) line emission.
\label{LCO2Md}}
\end{figure}

\clearpage

\begin{figure}
\plotone{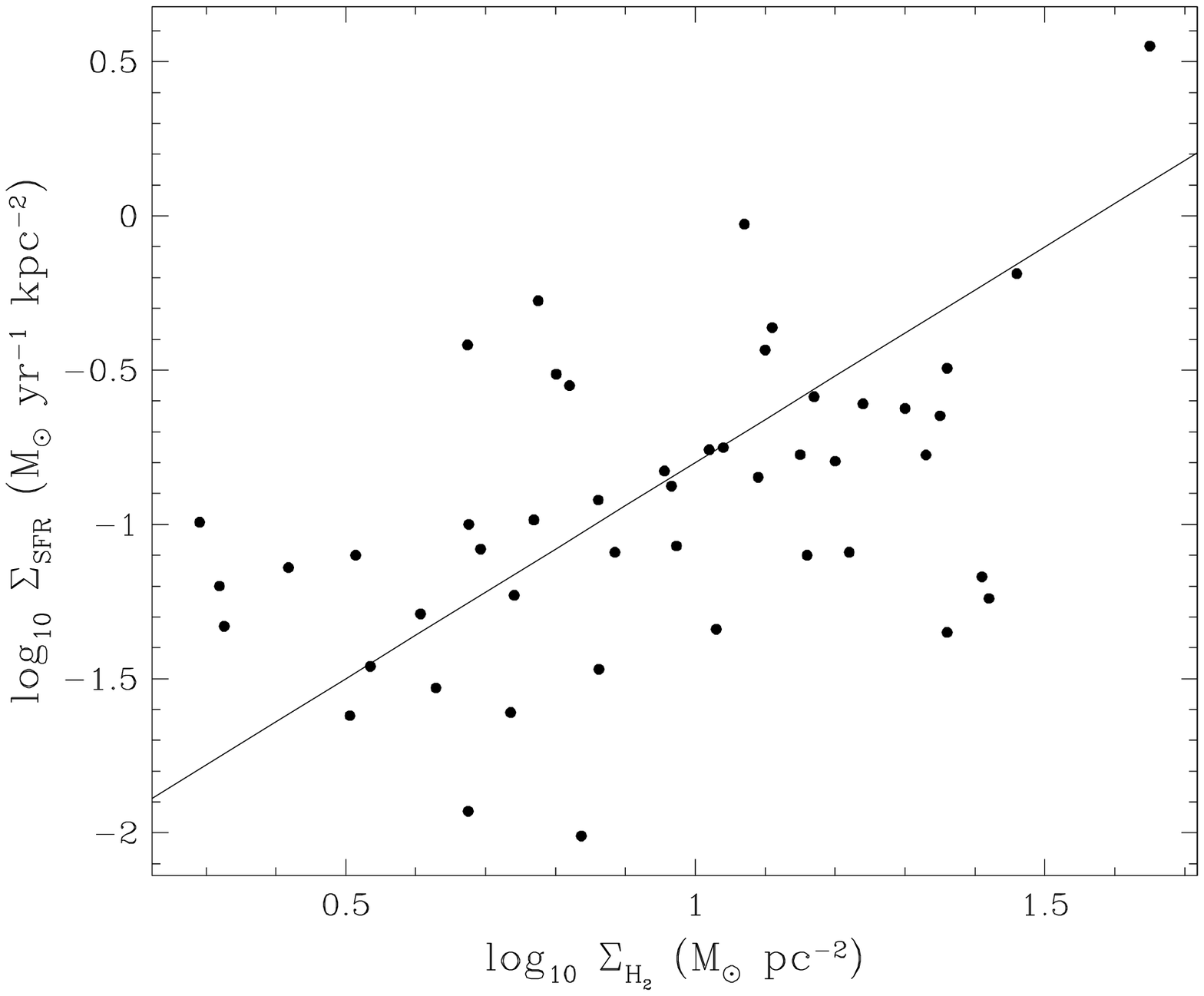}
\caption{Relation between the $SFR$ per unit area and the surface density of H$_2$ 
measured within a 15$^{\prime\prime}$ beam. 
\label{sfr2mh2}}
\end{figure}

\clearpage 

\begin{figure}
\plotone{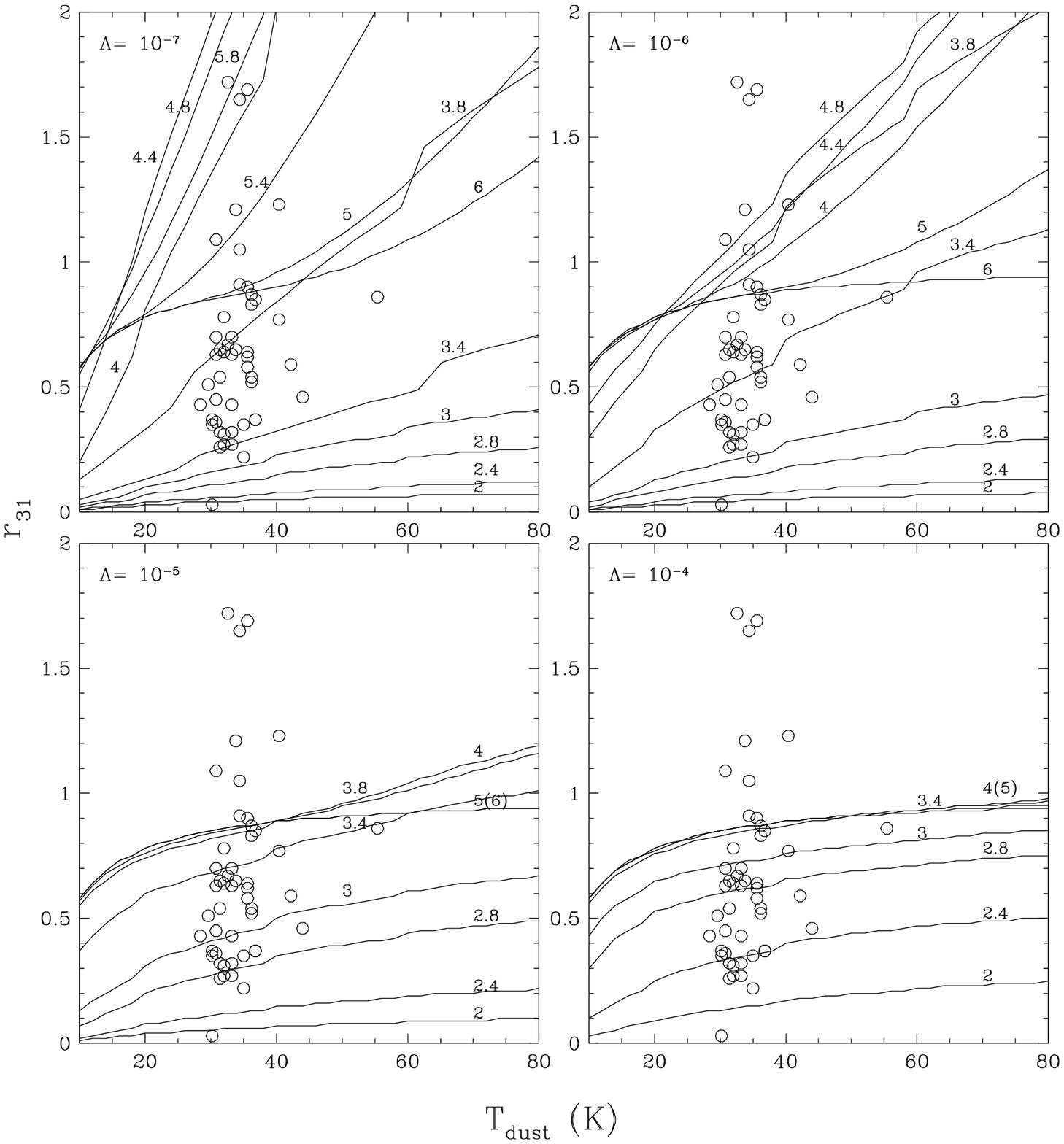}
\caption{Plots of $r_{31}$ versus dust temperature $T_{dust}$ for different densities 
of molecular hydrogen log$_{10}$$n$(H$_2$) based on LVG model computations for four 
$\Lambda$ values between 10$^{-7}$ and 10$^{-4}$ (km s$^{-1}$ pc$^{-1}$)$^{-1}$. 
Also shown are the observed data, assuming $T_{kin}$ = $T_{dust}$. 
The $^{12}$CO abundance $Z_{CO}$ is fixed at 10$^{-4}$, and the isotope abundance 
ratio $\Big[\frac{^{12}CO}{^{13}CO}\Big]$ = 40. 
\label{r2Td}}
\end{figure}

\clearpage 

\begin{figure}
\plotone{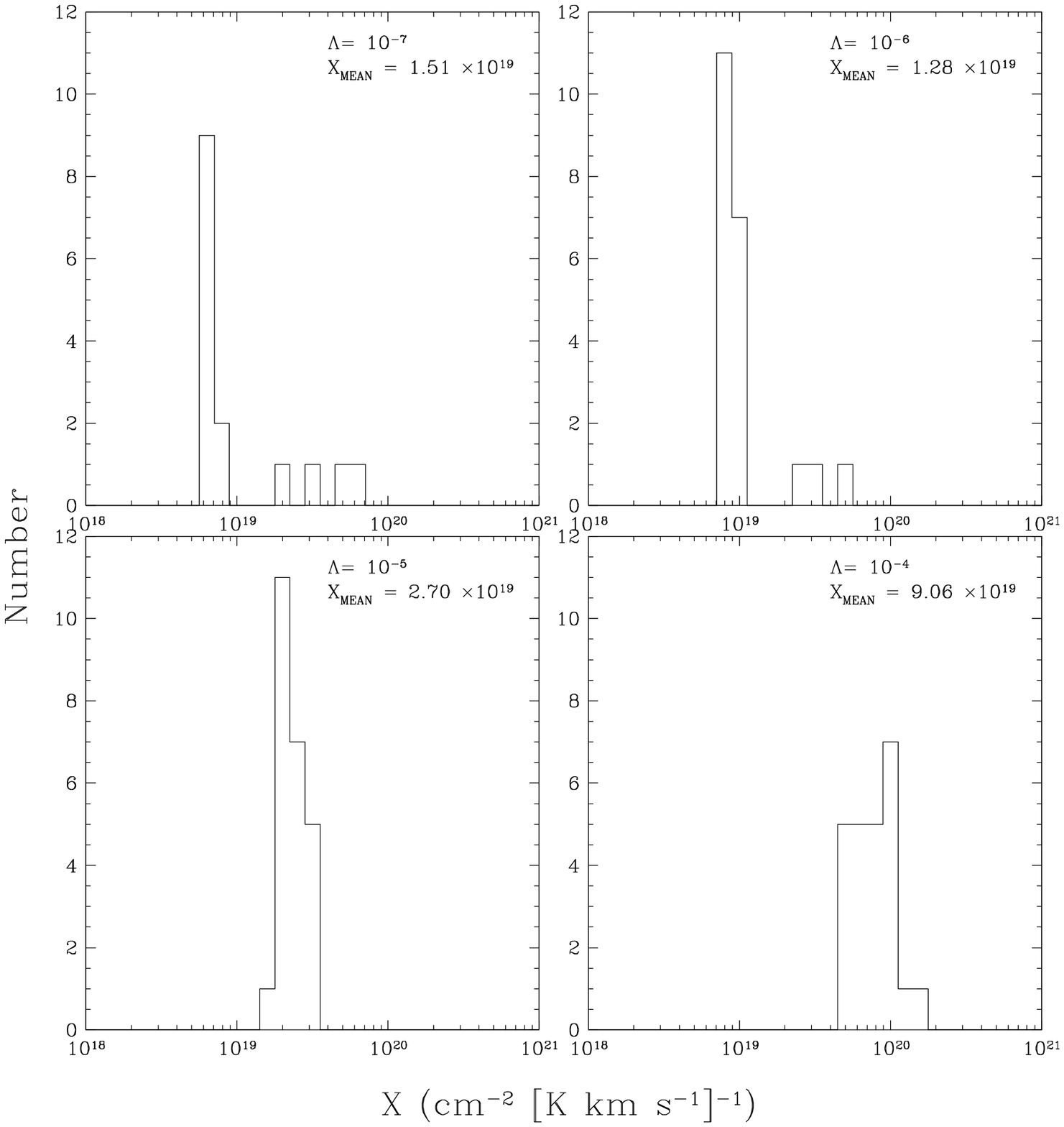}
\caption{Distribution of the $X$ conversion factor derived from LVG computations 
and its mean value $X_{MEAN}$ for different values of $\Lambda$ = 
$\frac{Z_{CO}}{(dv/dr)}$. The units of $X$ is cm$^{-2}$[K km s$^{-1}$]$^{-1}$ 
and $\Lambda$ is in (km s$^{-1}$ pc$^{-1}$)$^{-1}$. 
\label{Xhis}}
\end{figure}

\clearpage 

\begin{figure}
\plotone{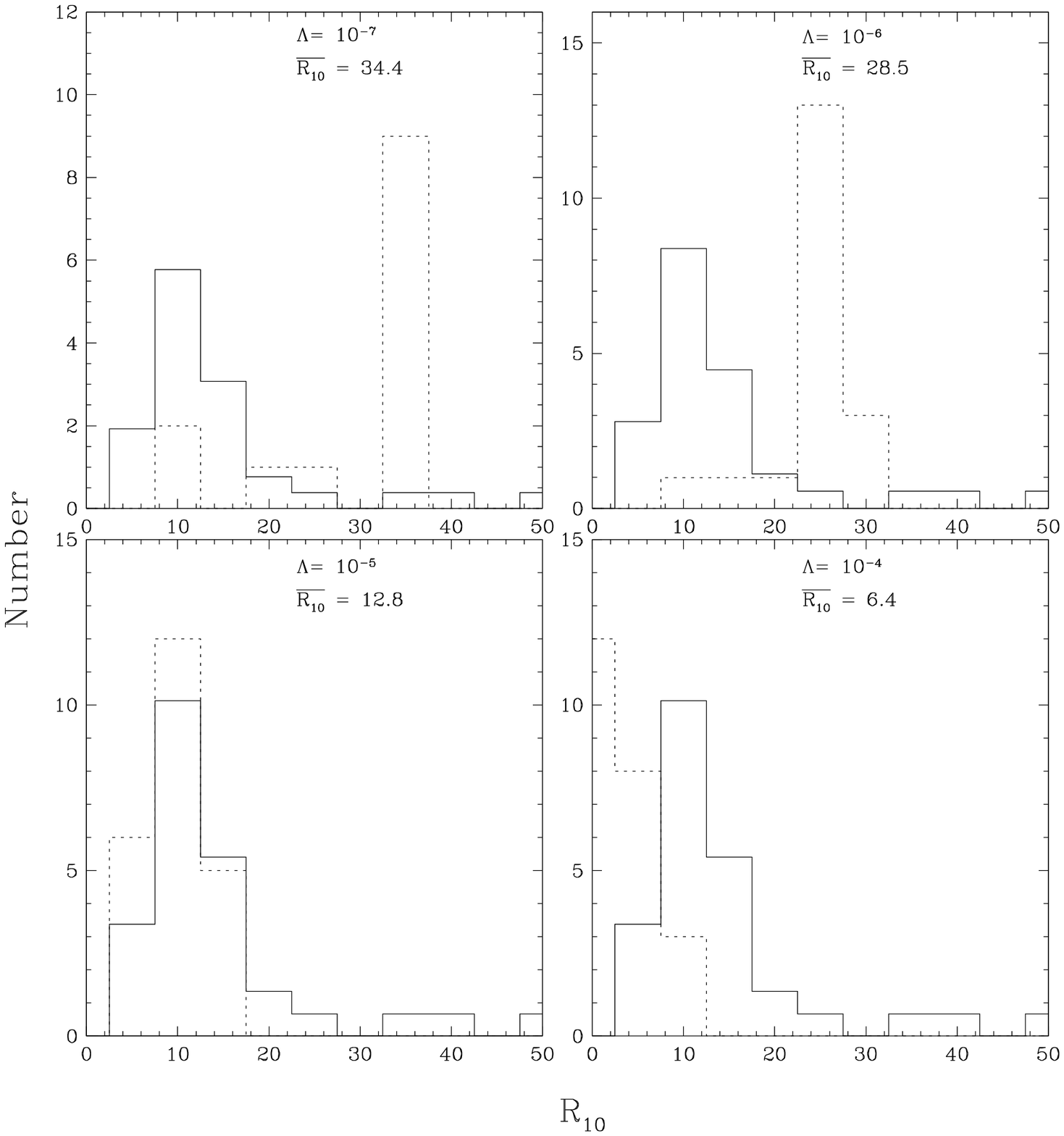}
\caption{Histograms of the $^{12}$CO(1-0)/$^{13}$CO(1-0) intensity ratio $R_{10}$ for 
different values of $\Lambda$. Dotted lines: LVG modeling results from this work. 
Solid lines: data taken from Aalto et al. (1995) and Taniguchi et al. (1999). The 
$\Big[\frac{^{12}CO}{^{13}CO}\Big]$ abundance ratio is fixed at 40. The values for 
$R_{10}$ observed correspond to a galaxy sample with $L_{FIR}$ $\ge$ 10$^{10}$ 
L$_{\odot}$, similar to the SLUGS subsample.
\label{R10his}}
\end{figure}

\clearpage

\begin{deluxetable}{lrrcrccccccc}
\tabletypesize{\tiny}
\tablecaption{$^{12}$CO line parameters of observed 60 IR luminous SLUGS galaxies. 
\label{tbl-1}}
\tablewidth{0pt}
\tablehead{
\colhead{Source\tablenotemark{a}} & \colhead{RA$_{2000}$\tablenotemark{b}} & \colhead{Dec$_{2000}$\tablenotemark{c}} & \colhead{$cz$\tablenotemark{d}} & \colhead{$D$\tablenotemark{e}} &  \colhead{$I_{10}$\tablenotemark{f}} & \colhead{$\pm\sigma {I_{10}}$\tablenotemark{g}} & \colhead{$I_{32}$\tablenotemark{h}} & \colhead{$\pm\sigma {I_{32}}$\tablenotemark{i}} & \colhead{$r_{31}$\tablenotemark{j}} & \colhead{-$\sigma_{31}$\tablenotemark{k}} & \colhead{+$\sigma_{31}$\tablenotemark{l}} \\
 & \colhead{h m s} & \colhead{$^{\circ}$ $^{\prime}$ $^{\prime}$$^{\prime}$} & \colhead{(km s$^{-1}$)} & \colhead{(Mpc)} & \colhead{(K km s$^{-1}$)} & \colhead{} & \colhead{(K km s$^{-1}$)} & \colhead{} & \colhead{} & \colhead{} & \colhead{}
}
\startdata
UGC 5376 &10 00 27.1 &+03 22 28 &2050 &27.2 &13.0 &3.3 &15.8 &3.3 &1.21 &-0.45 &0.33 \\
NGC 3094 &10 01 26.0 &+15 46 14 &2404 &31.9 &48.5 &10.1 &37.5 &7.7 &0.77 &-0.25 &0.20 \\
NGC 3110 &10 04 01.9 &-06 28 29 &5034 &66.3 &50.6 &10.7 &39.6 &8.2 &0.78 &-0.26 &0.21 \\
IR 1017+08 &10 19 59.9 &+08 13 34 &14390 &185.2 &16.1 &4.6 &7.4 &3.2 &0.46 &-0.28 &0.22 \\
NGC 3221 &10 22 20.0 &+21 34 10 &4110 &54.2 &33.5 &7.1 &16.9 &3.5 &0.51 &-0.17 &0.13 \\
NGC 3367 &10 46 34.8 &+13 45 02 &3037 &40.2 &38.0 &7.9 &12.0 &2.5 &0.32 &-0.11 &0.08 \\
IR 1056+24 &10 59 18.1 &+24 32 34 &12921 &166.9 &38.1 &8.0 &24.3 &5.3 &0.64 &-0.22 &0.17 \\
ARP 148 &11 03 53.9 &+40 51 00 &10350 &134.5 &8.8 &2.9 &5.5 &1.4 &0.62 &-0.29 &0.21 \\
NGC 3583 &11 14 10.8 &+48 19 03 &2136 &28.3 &44.0 &9.4 &18.7 &4.0 &0.43 &-0.14 &0.11 \\
MCG +00-29 &11 21 10.9 &-02 59 13 &7646 &100.0 &\nodata &\nodata &27.5 &5.8 &\nodata &\nodata &\nodata \\
UGC 6436 &11 25 45.0 &+14 40 36 &10243 &133.2 &53.3 &11.1 &18.4 &3.8 &0.35 &-0.11 &0.09 \\
NGC 3994p &11 57 36.8 &+32 16 39 &3118 &41.3 &21.2 &6.3 &14.9 &3.6 &0.70 &-0.31 &0.23 \\
NGC 3995p &11 57 44.1 &+32 17 40 &3254 &43.0 &23.2 &5.1 &6.3 &1.8 &0.27 &-0.11 &0.09 \\
NGC 4045 &12 02 42.3 &+01 58 38 &1981 &26.3 &57.3 &11.7 &14.8 &3.1 &0.26 &-0.08 &0.07 \\
IR 1211+03 &12 13 46.1 &+02 48 40 &21885 &276.7 &\nodata &\nodata &11.0 &3.0 &\nodata &\nodata &\nodata \\
NGC 4273 &12 19 56.0 &+05 20 34 &2378 &31.5 &32.9 &6.9 &21.5 &4.5 &0.65 &-0.22 &0.17 \\
IR 1222-06 &12 25 03.9 &-06 40 53 &7902 &103.3 &9.1 &3.6 &11.2 &2.5 &1.23 &-0.63 &0.44 \\
NGC 4418 &12 26 54.7 &-00 52 39 &2179 &28.9 &54.8 &11.5 &47.4 &9.7 &0.86 &-0.28 &0.22 \\
NGC 4433 &12 27 38.6 &-08 16 45 &3000 &39.7 &108.9 &22.7 &40.8 &8.4 &0.37 &-0.12 &0.10 \\
NGC 4793 &12 54 40.7 &+28 56 19 &2484 &32.9 &71.5 &14.5 &26.2 &5.3 &0.37 &-0.12 &0.09 \\
NGC 4922 &13 01 24.8 &+29 18 36 &7071 &92.6 &12.3 &5.1 &$<$ 2.04 &\nodata &\nodata &\nodata &\nodata \\
NGC 5020 &13 12 39.9 &+12 35 59 &3362 &44.5 &39.2 &8.0 &20.2 &4.3 &0.52 &-0.23 &0.18 \\
IC 860 &13 15 03.5 &+24 37 08 &3347 &44.3 &$<$ 5.07 &\nodata &$<$ 2.67 &\nodata &\nodata &\nodata &\nodata \\
UGC 8387 &13 20 35.3 &+34 08 22 &7000 &91.7 &90.6 &18.4 &63.3 &12.9 &0.70 &-0.21 &0.16 \\
NGC 5104 &13 21 23.1 &+00 20 32 &5578 &73.4 &62.6 &12.7 &33.6 &6.8 &0.54 &-0.15 &0.12 \\
NGC 5256 &13 38 17.5 &+12 35 58 &8353 &109.1 &\nodata &\nodata &12.3 &2.9 &\nodata &\nodata &\nodata \\
NGC 5257p &13 39 53.0 &+00 50 22 &6757 &88.6 &33.1 &7.1 &20.7 &4.3 &0.63 &-0.35 &0.28 \\
NGC 5258p &13 39 57.9 &+00 49 58 &6798 &89.1 &37.7 &8.4 &16.1 &3.5 &0.43 &-0.12 &0.09 \\
UGC 8739 &13 49 14.2 &+35 15 23 &5032 &66.3 &44.2 &9.0 &48.3 &9.9 &1.09 &-0.05 &0.05 \\
NGC 5371 &13 55 39.0 &+40 27 31 &2553 &33.8 &\nodata &\nodata &15.5 &3.3 &\nodata &\nodata &\nodata \\
NGC 5394p &13 58 33.6 &+37 27 13 &3472 &45.9 &101.5 &21.1 &36.0 &7.3 &0.35 &-0.63 &0.44 \\
NGC 5395p &13 58 37.6 &+37 25 32 &3491 &46.1 &30.4 &6.6 &\nodata &\nodata &\nodata &\nodata &\nodata \\
NGC 5433 &14 02 36.0 &+32 30 38 &4354 &57.4 &24.1 &7.2 &40.6 &8.6 &1.69 &-0.27 &0.20 \\
NGC 5426p &14 03 25.1 &-06 04 09 &2621 &34.7 &18.2 &4.8 &5.0 &1.7 &0.27 &-0.31 &0.24 \\
NGC 5427p &14 03 25.6 &-06 01 42 &2678 &35.5 &12.1 &3.1 &7.7 &2.0 &0.64 &-0.26 &0.19 \\
ZW 247.020 &14 19 43.2 &+49 14 12 &7666 &100.3 &28.5 &6.2 &25.9 &5.5 &0.91 &-0.20 &0.16 \\
NGC 5600 &14 23 49.4 &+14 38 21 &2319 &30.7 &11.0 &3.0 &7.2 &1.5 &0.65 &-0.07 &0.06 \\
NGC 5653 &14 30 10.4 &+31 12 54 &3562 &47.1 &61.1 &12.6 &38.4 &7.7 &0.63 &-0.61 &0.45 \\
NGC 5665 &14 32 25.8 &+08 04 45 &2228 &29.5 &78.0 &16.6 &16.8 &3.5 &0.22 &-0.30 &0.23 \\
NGC 5676 &14 32 46.7 &+49 27 29 &2114 &28.0 &\nodata &\nodata &14.7 &3.2 &\nodata &\nodata &\nodata \\
NGC 5713 &14 40 11.3 &-00 17 27 &1900 &25.2 &16.4 &4.1 &27.2 &5.5 &1.65 &-0.29 &0.22 \\
UGC 9618p &14 57 00.5 &+24 36 42 &9900 &128.8 &\nodata &\nodata &5.2 &1.7 &\nodata &\nodata &\nodata \\
NGC 5792 &14 58 22.8 &-01 05 27 &1924 &25.5 &35.0 &7.4 &31.6 &6.5 &0.90 &-0.15 &0.11 \\
ZW 049.057 &15 13 13.1 &+07 13 27 &3897 &51.5 &27.4 &6.0 &22.8 &4.9 &0.83 &-0.57 &0.42 \\
NGC 5900 &15 15 05.0 &+42 12 32 &2511 &33.3 &74.8 &15.4 &33.6 &6.9 &0.45 &-0.22 &0.16 \\
1 ZW 107 &15 18 06.1 &+42 44 45 &12039 &155.8 &\nodata &\nodata &14.6 &3.3 &\nodata &\nodata &\nodata \\
NGC 5929p &15 26 06.1 &+41 40 14 &2492 &33.0 &5.9 &3.4 &5.1 &1.2 &0.87 &-0.22 &0.17 \\
NGC 5930p &15 26 07.9 &+41 40 34 &2672 &35.4 &21.5 &5.7 &11.7 &2.6 &0.54 &-0.60 &0.46 \\
IR 1525+36 &15 26 59.4 &+35 58 37 &16602 &212.6 &\nodata &\nodata &10.3 &3.3 &\nodata &\nodata &\nodata \\
NGC 5936 &15 30 00.8 &+12 59 21 &4004 &52.9 &56.4 &11.8 &37.8 &7.6 &0.67 &-0.28 &0.22 \\
NGC 5937 &15 30 46.2 &-02 49 45 &2807 &37.2 &18.8 &4.3 &32.2 &6.5 &1.72 &-0.13 &0.10 \\
NGC 5953p &15 34 32.3 &+15 11 38 &1965 &26.1 &45.7 &9.8 &38.7 &7.9 &0.85 &-0.19 &0.15 \\
NGC 5954p &15 34 35.0 &+15 12 00 &1959 &26.0 &38.6 &8.3 &14.4 &3.2 &0.37 &-0.10 &0.08 \\
ARP 220 &15 34 57.1 &+23 30 11 &5434 &71.5 &126.3 &26.2 &74.2 &14.9 &0.59 &-0.11 &0.09 \\
IR 1533-05 &15 36 11.7 &-05 23 52 &8186 &107.0 &\nodata &\nodata &8.9 &2.2 &\nodata &\nodata &\nodata \\
NGC 5962 &15 36 31.7 &+16 36 29 &1958 &26.0 &86.2 &17.6 &26.9 &5.5 &0.31 &-0.22 &0.16 \\
NGC 5990 &15 46 16.5 &+02 24 56 &3839 &50.7 &86.7 &20.0 &28.1 &5.9 &0.32 &-0.38 &0.29 \\
NGC 6052 &16 05 12.9 &+20 32 32 &4716 &62.1 &30.6 &8.0 &17.8 &3.7 &0.58 &-0.12 &0.09 \\
MCG +01-42 &16 30 56.5 &+04 04 59 &7342 &96.1 &26.4 &6.3 &27.6 &5.7 &1.05 &-0.37 &0.28 \\
NGC 6181 &16 32 21.0 &+19 49 36 &2375 &31.5 &60.0 &12.8 &21.3 &4.4 &0.36 &-0.12 &0.09 \\ \hline
\enddata

\tablenotetext{a}{Galaxy name, the letter p denotes those galaxies that belong to a pair.}
\tablenotetext{b}{Right ascension J2000 epoch from NED}
\tablenotetext{c}{Declination J2000 epoch from NED.}
\tablenotetext{d}{Recession velocity from NED.}
\tablenotetext{e}{Galaxy proper distance to the object at the present time.}
\tablenotetext{f}{The $^{12}$CO(1-0) line intensity $I_{10}$ = $\int T_{mb}dv$ in K km s$^{-1}$. Upper limits refer to 3 times the standard error due to the noise.}
\tablenotetext{g}{Standard error of $I_{10}$.}
\tablenotetext{h}{The $^{12}$CO(3-2) line intensity $I_{32}$ = $\int T_{mb}dv$ in K km s$^{-1}$. Upper limits refer to 3 times the standard error due to the noise.}
\tablenotetext{i}{Standard error of $I_{32}$.}
\tablenotetext{j}{The line intensity ratio $r_{31}$.}
\tablenotetext{k}{The negative uncertainty in $r_{31}$.}
\tablenotetext{l}{The positive uncertainty in $r_{31}$.}

\end{deluxetable}

\clearpage

\begin{deluxetable}{lccccc}
\tabletypesize{\scriptsize}
\tablecaption{CO luminosities and masses. 
\label{tbl-2}}
\tablewidth{0pt}
\tablehead{
\colhead{Source\tablenotemark{a}} & \colhead{P.B.\tablenotemark{b}} & \colhead{log$_{10}$$L_{^{12}CO(1-0)}$\tablenotemark{c}} & \colhead{log$_{10}$$L_{^{12}CO(3-2)}$\tablenotemark{d}} & \colhead{log$_{10}$$M$(H$_2$)\tablenotemark{e}} & \colhead{$SFE$\tablenotemark{f}} \\
 & \colhead{(kpc)} & \colhead{(K km s$^{-1}$ Mpc$^2$ $\Omega_b$)} & \colhead{(K km s$^{-1}$ Mpc$^2$ $\Omega_b$)} & \colhead{(M$_\odot$)} & \colhead{(L$_\odot$/M$_\odot$)}  
}
\startdata
UGC 5376 &1.99 &3.98 &4.06 &7.34 &177 \\
NGC 3094 &2.34 &4.69 &4.58 &8.05 &102 \\
NGC 3110 &4.90 &5.34 &5.23 &8.71 &98 \\
IR 1017+08 &14.12 &5.72 &5.39 &9.12 &96 \\
NGC 3221 &4.00 &4.99 &4.69 &8.36 &43 \\
NGC 3367 &2.95 &4.78 &4.28 &8.15 &32 \\
IR 1056+24 &12.66 &6.01 &5.81 &9.40 &197 \\
ARP 148 &10.12 &5.19 &4.98 &8.57 &302 \\
NGC 3583 &2.07 &4.54 &4.17 &7.91 &68 \\
MCG +00-29 &7.46 &\nodata &5.43 &\nodata &\nodata \\
UGC 6436 &10.02 &5.96 &5.50 &9.35 &94 \\
NGC 3994p &3.03 &4.55 &4.40 &7.92 &61 \\
NGC 3995p &3.16 &4.63 &4.07 &7.99 &59 \\
NGC 4045 &1.92 &4.59 &4.01 &7.96 &82 \\
IR 1211+03 &21.59 &\nodata &5.89 &\nodata &\nodata \\
NGC 4273 &2.31 &4.51 &4.33 &7.87 &84 \\
IR 1222-06 &7.71 &4.98 &5.07 &8.36 &142 \\
NGC 4418 &2.12 &4.66 &4.59 &8.02 &461 \\
NGC 4433 &2.92 &5.23 &4.80 &8.60 &46 \\
NGC 4793 &2.41 &4.89 &4.45 &8.25 &28 \\
NGC 4922 &6.90 &5.01 &\nodata &8.39 &40 \\
NGC 5020 &3.27 &4.88 &4.60 &8.25 &68 \\
IC 860 &3.25 &\nodata &\nodata &\nodata &\nodata \\
UGC 8387 &6.83 &5.87 &5.72 &9.25 &130 \\
NGC 5104 &5.43 &5.52 &5.25 &8.89 &96 \\
NGC 5256 &8.15 &\nodata &5.16 &\nodata &\nodata \\
NGC 5257p &6.59 &5.41 &5.20 &8.78 &27 \\
NGC 5258p &6.63 &5.47 &5.10 &8.84 &24 \\
UGC 8739 &4.90 &5.28 &5.32 &8.65 &73 \\
NGC 5371 &2.48 &\nodata &4.25 &\nodata &\nodata \\
NGC 5394p &3.38 &5.32 &4.87 &8.69 &15 \\
NGC 5395p &3.39 &4.81 &\nodata &8.17 &8 \\
NGC 5433 &4.24 &4.89 &5.12 &8.26 &160 \\
NGC 5426p &2.55 &4.34 &3.77 &7.70 &14 \\
NGC 5427p &2.60 &4.18 &3.98 &\nodata &\nodata \\
ZW 247.020 &7.48 &5.45 &5.40 &8.82 &169 \\
NGC 5600 &2.25 &4.01 &3.83 &7.37 &121 \\
NGC 5653 &3.46 &5.13 &4.92 &8.49 &68 \\
NGC 5665 &2.16 &4.83 &4.16 &8.19 &11 \\
NGC 5676 &2.05 &\nodata &4.06 &\nodata &\nodata \\
NGC 5713 &1.85 &4.02 &4.24 &7.38 &282 \\
UGC 9618p &9.68 &\nodata &4.92 &\nodata &\nodata \\
NGC 5792 &1.87 &4.36 &4.31 &7.72 &127 \\
ZW 049.057 &3.79 &4.85 &4.77 &8.22 &517 \\
NGC 5900 &2.44 &4.91 &4.57 &8.28 &53 \\
1 ZW 107 &11.79 &\nodata &5.53 &\nodata &\nodata \\
NGC 5929p &2.42 &3.80 &3.74 &\nodata &\nodata \\
NGC 5930p &2.60 &4.43 &4.16 &7.79 &99 \\
IR 1525+36 &16.31 &\nodata &5.64 &\nodata &\nodata \\
NGC 5936 &3.90 &5.19 &5.02 &8.56 &82 \\
NGC 5937 &2.73 &4.41 &4.64 &7.77 &469 \\
NGC 5953p &1.91 &4.49 &4.42 &7.85 &58 \\
NGC 5954p &1.90 &4.41 &3.99 &7.78 &61 \\
ARP 220 &5.29 &5.80 &5.57 &9.17 &467 \\
IR 1533-05 &7.99 &\nodata &5.00 &\nodata &\nodata \\
NGC 5962 &1.90 &4.76 &4.26 &8.12 &13 \\
NGC 5990 &3.73 &5.34 &4.85 &8.71 &58 \\
NGC 6052 &4.59 &5.07 &4.83 &8.44 &102 \\
MCG +01-42 &7.16 &5.38 &5.40 &8.75 &247 \\
NGC 6181 &2.31 &4.77 &4.32 &8.13 &26 \\ \hline
\enddata

\tablenotetext{a}{Galaxy name, the letter p denotes those galaxies that belong to a pair.}
\tablenotetext{b}{The projected beamsize (P.B.) at the source.}
\tablenotetext{c}{The $^{12}$CO(1-0) luminosity measured within a 15$^{\prime\prime}$ 
beam, $\Omega_b$ = 1 beam area.}
\tablenotetext{d}{The $^{12}$CO(3-2) luminosity measured within a 15$^{\prime\prime}$ 
beam, $\Omega_b$ = 1 beam area.}
\tablenotetext{e}{Molecular hydrogen gas mass measured within a 15$^{\prime\prime}$ 
beam.}
\tablenotetext{f}{Star formation efficiency measured within a 15$^{\prime\prime}$ 
beam.}

\end{deluxetable}

\end{document}